\documentclass[12pt]{article}
\usepackage[dvips]{epsfig}
\usepackage{latexsym}
\usepackage[bf,footnotesize]{caption}	
\begin{document}
\setlength{\unitlength}{1mm}

\newcommand{\ba} {\begin{eqnarray}}
\newcommand{\ea} {\end{eqnarray}}
\newcommand{\be}{\begin{equation}}
\newcommand{\ee}{\end{equation}}
\newcommand{\n}[1]{\label{#1}}
\newcommand{\eq}[1]{Eq.(\ref{#1})}
\newcommand{\ind}[1]{\mbox{\tiny{#1}}}
\renewcommand\theequation{\thesection.\arabic{equation}}

\newcommand{\nn}{\nonumber \\ \nonumber \\}
\newcommand{\nl}{\\  \nonumber \\}
\newcommand{\pr}{\partial}
\renewcommand{\vec}[1]{\mbox{\boldmath$#1$}}

\title{{\hfill {\small Alberta-Thy-08-99 } } \vspace*{2cm} \\
Quantum Effects in the Presence of Expanding Semi-Transparent
Spherical Mirrors}
\author{\\
V. Frolov\thanks{e-mail: frolov@phys.ualberta.ca}
\  and 
D. Singh\thanks{e-mail: 
singh@phys.ualberta.ca}
\date{\today}}
\maketitle
\noindent 
{
\centerline{ \em
Theoretical Physics Institute, Department of Physics,} \\ 
\centerline{ \em University of Alberta, Edmonton, Canada T6G 2J1}
}
\bigskip

\begin{abstract}
We study quantum effects in the presence of a spherical semi-transparent
mirror or a system of two concentric mirrors which expand with a
constant acceleration in a flat $D$-dimensional spacetime. Using the
Euclidean approach, we obtain expressions for $\langle \hat{\varphi}^2
\rangle^{\ind{ren}}$ and $\langle \hat{T}^{\mu}_{\nu}
\rangle^{\ind{ren}}$ for a scalar non-minimally coupled massless field.
Explicit expressions are obtained for $\langle \hat{\varphi}^2
\rangle^{\ind{ren}}$  and the energy fluxes at ${\cal J}^{\pm}$ generated by
such mirrors in the physical spacetime and their properties are
discussed.
\end{abstract}

\bigskip

{\it PACS number(s): 03.70.+k, 11.10.-z, 42.50.Lc}

\baselineskip=.6cm

\newpage

\section{Introduction}

One of the simplest examples of  a general problem of a quantum 
theory in a given external field is the
study of quantum effects in the presence of mirror-like boundaries.
A mirror can be considered as a (finite
or infinite) potential barrier, the position of which is sharply
localized in space. If mirror-like boundaries are static, the ground
state of the system is still a vacuum, but the energy of this state is
different from the energy of the vacuum in the absence of boundaries.
This is the well-known  {\em Casimir effect}
\cite{Casi:48}--\cite{Most:97}.  A new phenomenon, the {\em dynamical
Casimir effect} \cite{DeWi:75}-\cite{BiDa:82}, occurs for moving
boundaries.  In the general case, a motion of mirror-like boundaries 
creates energy fluxes and particles from the vacuum.
Non-uniformly moving mirrors radiate ``photon'' pairs. In many
aspects,  this effect  is similar to the effect of parametric
excitation of a quantum oscillator.

Since the pioneering work of Moore \cite{Moor:70} who studied the
creation of photons by moving mirrors for a 1-dimensional cavity, 
numerous publications investigating different aspects of the dynamical
Casimir effect  have appeared. Fulling and Davies \cite{FuDa:76} found
that the back reaction force generated by emission of photons in
2-dimensional spacetime (that is by a 1-dimensional mirror) is
proportional to the third derivative of the mirror displacement. 
Two-dimensional problems for conformal-invariant fields  allow quite a
complete study because of the presence of an infinite-dimensional group
of conformal transformations. One can use these transformations to map
the original problem with dynamical boundaries to a corresponding 
static problem which allows a solution. 

Absence of such a wide group of invariance
makes the study of problems in higher dimensions much more
complicated. Simple case when plane boundaries are moving with constant
velocities was considered in \cite{BoPeRo:84,BoPeRo:86}. 
Perturbation methods were used for study of physically
interesting case of  the emission of photons by a vibrating cavity in a
realistic $(3+1)$-dimensional spacetime \cite{Calu:92}-\cite{CoKa:98}.
This study demonstrates that the mirror-induced radiation, which is
usually  far too weak to be observed for speeds much below the speed of
light,  can be enhanced by orders of magnitudes by resonance effect due
to the cavity finesse. A review of results for radiation from moving
mirrors and bodies with finite refractive index can be found in
\cite{BaEb:93}--\cite{BaNo:96}.

Not so much work has been done in the study of radiation created by a
relativistically moving mirror in four (and higher) dimensional
spacetimes. Candelas and Deutsch \cite{CaDe:77} considered radiation
from a uniformly accelerated ideally-reflecting plane. The  method of
images was used in \cite{FrSe:79} and \cite{FrSe:80} to obtain the
exact Green function and to calculate the stress-energy produced by a
spherical ideal mirror, which is expanding in the four-dimensional
spacetime with a constant acceleration for a scalar and electromagnetic
fields, respectively. The latter calculations can be interesting in
connection with the study of quantum effects produced by relativistic bubbles
generated in the cosmological phase transitions (see e.g.
\cite{Cole:80} -- \cite{BeKuTk:87}). 

In this paper, we continue the study of quantum effects produced by spherical
mirrors expanding with a uniform acceleration. More concretely, we study
a scalar massless field propagating in a spacetime 
with partially transparent mirrors\footnote{For discussion of the
Casimir effect for a system with partially transparent boundaries see,
e.g. recent papers \cite{BoKiVa:99,BoVa:99} and references therein.}. 
The latter is described by $\delta$-like potentials. We consider two
problems: Problem $A$ for which the mirror potential is localized at
the hyperboloid $R^2-T^2=a^{-2}$, where $a$ is the value of the
acceleration and $R$ is the radius of the mirror at time $T$; and
Problem $B$, where there exist two expanding\footnote{Strictly speaking
for the motion with a
constant acceleration, a spherical mirror at first contracts from an
infinite size to a finite radius, where its velocity vanishes and only
after this it starts its expansion, so that the velocity of this
expansion reaches its asymptotic value $c$.
} concentric mirrors.  This allows us to demonstrate how partial
transparency of mirrors (which is a common feature of more realistic
models) modifies the expression for the radiation emitted by expanding
mirrors. After performing a Wick's rotation $T\rightarrow iX_0$, the
problem is reduced to finding a potential generated by a point-like
source in 4-dimensional space in the presence of one or two concentric
$\delta$-like spherical potentials. The latter problem can be solved by
decomposing into spherical harmonics. Since the solution of the problem
can be easily obtained in an arbitrary number of dimensions, we perform
the calculations for the general case of $D$-dimensional spacetime. 

The paper is organized as follows. Section 2 discusses the formulation
of the problem and considers two special models: Model $A$ with one
expanding partially transparent mirror, and Model $B$ with two
expanding concentric mirrors. Section 3 contains the calculation of the
Euclidean Green functions for both problems. The expressions for
$\langle \hat{\varphi}^2\rangle^{\ind{ren}}$ and $\langle
\hat{T}^{\mu}_{\nu}\rangle^{\ind{ren}}$ for the Euclidean formulation
are obtained in Sections 4 and 5, respectively. Radiation from a
spherical mirror, which expands with a constant acceleration and a
couple of expanding concentric mirrors, is calculated in Section 6.
Section 7 contains a discussion of the obtained results.

\section{Formulation of the Problem}\label{s2}
\setcounter{equation}0
\subsection{General formulas}
Our aim is to study the influence of an accelerated motion of
semi-transparent mirrors  on quantum fields. In our consideration, the
background geometry is flat. Nevertheless,   it is convenient to first
consider the general action for a scalar massless field in
curved  geometry
\be
S[\varphi ] = -\,\frac{1}{2}\int dx^D\,g^{1/2}\,\left(g^{\mu\nu}\,\varphi_{\!,\mu}\,
\varphi_{\!,\nu} + \xi R\,\varphi^2\right).  \n{2.1}
\ee
We keep the number of spacetime dimensions arbitrary and denote it by
$D$. We also denote by $d=D-1$ the number of spatial dimensions.
The field  equation is
\be
\Box \varphi -  \xi R\,\varphi = 0\,,  \n{2.2}
\ee
where $\Box = g^{-1/2}\,\partial_{\mu}\!\left(g^{1/2}\,g^{\mu\nu}\partial_{\nu}\right)$. The 
stress-energy tensor, $T_{\mu\nu}$, for the field has the form
\[
T_{\mu\nu} = (1-2\xi )\,\varphi_{\!,\mu}\,\varphi_{\!,\nu} + \left(2\xi - \frac{1}{2}\right)
g_{\mu\nu}\,g^{\alpha\beta}\,\varphi_{\!,\alpha}\,\varphi_{\!,\beta }
\]
\be
\hspace{2.9cm} - \,2\xi\varphi \left(\varphi_{;\mu\nu} - g_{\mu\nu}\,\varphi_{;\alpha}\,
\varphi^{;\alpha}\right) + \xi \left(R_{\mu\nu} - \frac{1}{2}g_{\mu\nu}\,R\right)
\varphi^2.  \n{2.3}
\ee
When the parameter $\xi$ of non-minimal coupling vanishes, $T_{\mu\nu}$
reduces to  the canonical stress-energy tensor. The case
\be
\xi = \frac{1}{4}\,\frac{D-2}{D-1}\equiv {d-1 \over 2d}  \n{2.4}
\ee
corresponds to the conformally invariant theory with $T_{\mu}^{\mu} = 0 $.

Consider now the quantum field $\hat{\varphi}$ in a flat spacetime and denote 
by $G^1(x,x')$ its Hadamard function
\be
G^1(x,x') = \langle \hat{\varphi}(x)\,\hat{\varphi}(x') + \hat{\varphi}(x')\,
\hat{\varphi}(x)\rangle \,.  \n{2.5}
\ee
Then for a quantum average of stress-energy tensor we have
\be
\langle T_{\mu\nu }(x)\rangle = \lim_{x'\rightarrow x}D_{\mu\nu '}\,
G^{(1)}(x,x')\,,  \n{2.6}
\ee
where
\[
D_{\mu\nu '} = \left(\frac{1}{2}-\xi\right)\nabla_{\!\mu}\,\nabla_{\!\nu'} + 
\left(\xi -\frac{1}{4}\right)g_{\mu\nu '}\,g^{\alpha\beta '}\nabla_{\!\alpha}\,
\nabla_{\!\beta '}
\]
\be
\hspace{-1.6cm} -\,\frac{1}{2}\,\xi\left(\nabla_{\!\mu}\,\nabla_{\!\nu} + 
\nabla_{\!\mu'}\,\nabla_{\!\nu'}\right).  \n{2.7}
\ee
Here $\nabla_{\!\mu}$ and $\nabla_{\!\mu'}$ are gradient operators acting 
on the first and second arguments of $G(x,x')$, respectively.

We write the metric of the flat spacetime metric in Cartesian coordinates as
\be
ds^2 = -\,dT^2 + \sum_{i=1}^{d}(dX_i)^2\,.  \n{2.8}
\ee
We shall also use the Euclidean metric obtained from 
(\ref{2.8}) by a Wick's rotation $T\rightarrow iX_0$
\be
ds^2_{E} =  \sum_{\mu =0}^{d}(dX_\mu)^2\,.  \n{2.9}
\ee

The Green function $G(x, x')$ of the quantum field $\hat{\varphi}$ in
the  flat spacetime is defined as the solution of the equation
\be
\Box G(x,x') = - \delta (x,x'),   \n{2.10}
\ee
obeying given boundary conditions. In the absence of boundaries, the 
Euclidean Green function $G_0$ is singled out by the condition that it
vanishes at infinity and is written in the Cartesian coordinates 
(\ref{2.9}) as
\be
G_0(x,x') = \frac{\Gamma\left(\frac{D}{2}-1\right)}{4\pi^{D/2}}\,
\frac{1}{|X-X'|^{D-2}}\,,  \n{2.11}
\ee
where
\be
|X-X'|^{2} = \sum_{\mu =0}^d(X_\mu-X'_\mu)^2\,.  \n{2.12}
\ee
This Green function is a solution of equation (\ref{2.10}) with
\be
\Box _E = \sum_{\mu =0}^d\frac{\partial^2}{\partial X_\mu^2}\,.  \n{2.13}
\ee

The renormalized value of the stress-energy tensor is defined by 
(\ref{2.6}), where instead of $G^{(1)}(x,x'),$ we must substitute
\be
G^{(1)}_{\ind{ren}}(x,x') = G^{(1)}(x,x') - 2G_{0}(x,x')\,.  \n{2.14}
\ee
Similarly, for $\langle\hat{\varphi}^2\rangle$ describing fluctuations
of the field  $\hat{\varphi},$ we have
\be
\langle\hat{\varphi}^2(x)\rangle^{\ind{ren}} ={1\over 2} \lim_{x'\rightarrow x}
G^{(1)}_{\ind{ren}}(x,x') \,.  \n{2.15}
\ee

\subsection{Models}

To describe a semi-transparent mirror, we include in the equation
(\ref{2.10}) 
an additional term, $-V_{\Sigma}\, \hat{\varphi}$, with a potential 
$V_{\Sigma}$ of the following form:
\be
V_{\Sigma} = V_0\,\delta(\Sigma )\,,   \n{2.16}
\ee
where $\Sigma$ is a timelike $d$-dimensional hypersurface representing 
the motion of a $(d-1)$-dimensional mirror surface. For finite values of
$V_0,$ a  mirror can be partially penetrated by the field
$\hat{\varphi}$. In the limit  $V_0\rightarrow\infty$ the penetration
coefficient vanishes, and one has  an ideal mirror which reflects all
the frequencies.

We consider two models  of moving semi-transparent mirrors.

\subsubsection{Model $A$}

\begin{figure}[t]
\hfill
{\epsfig{file=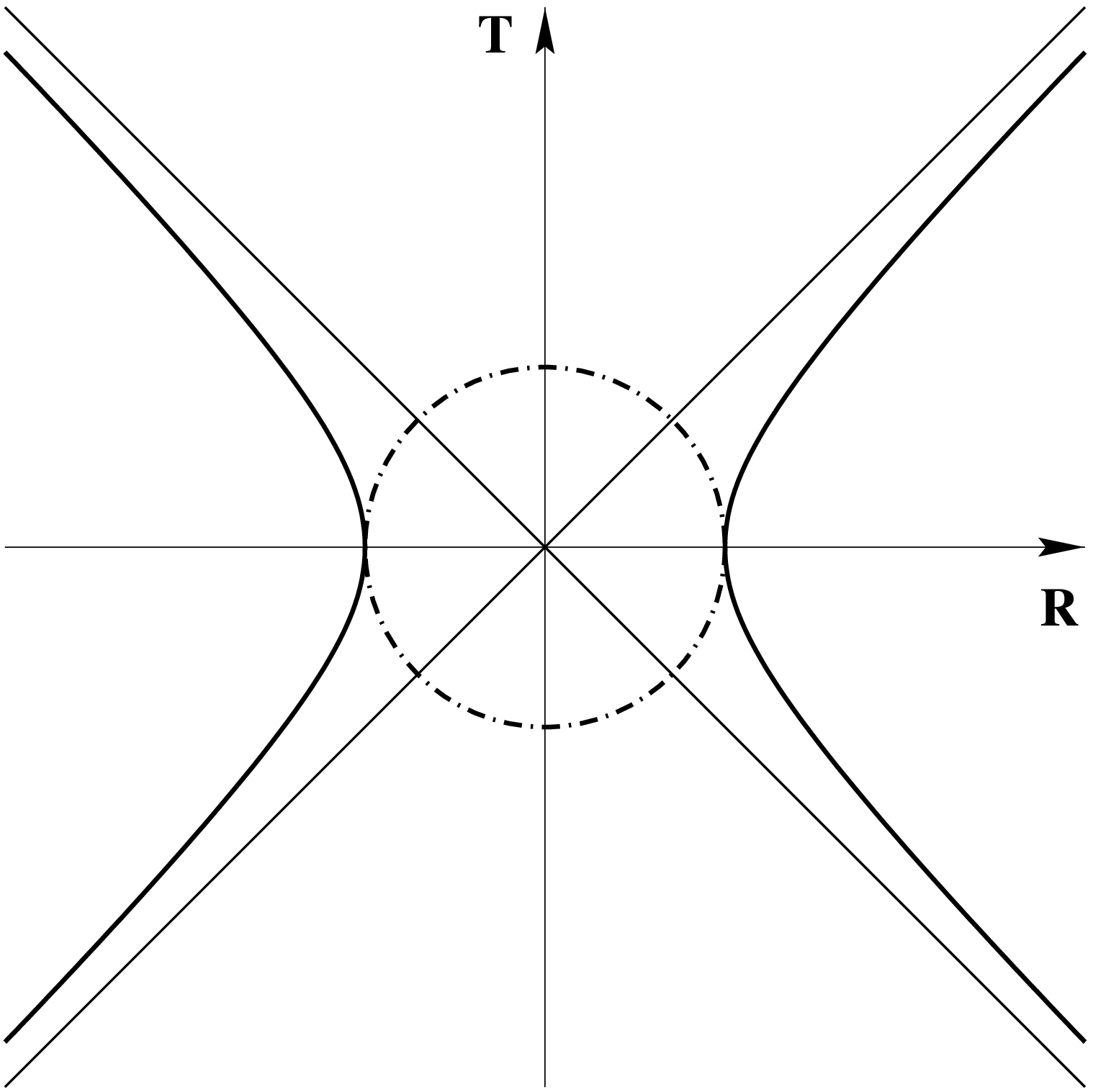, width=6.0cm}}
\hfill
{\epsfig{file=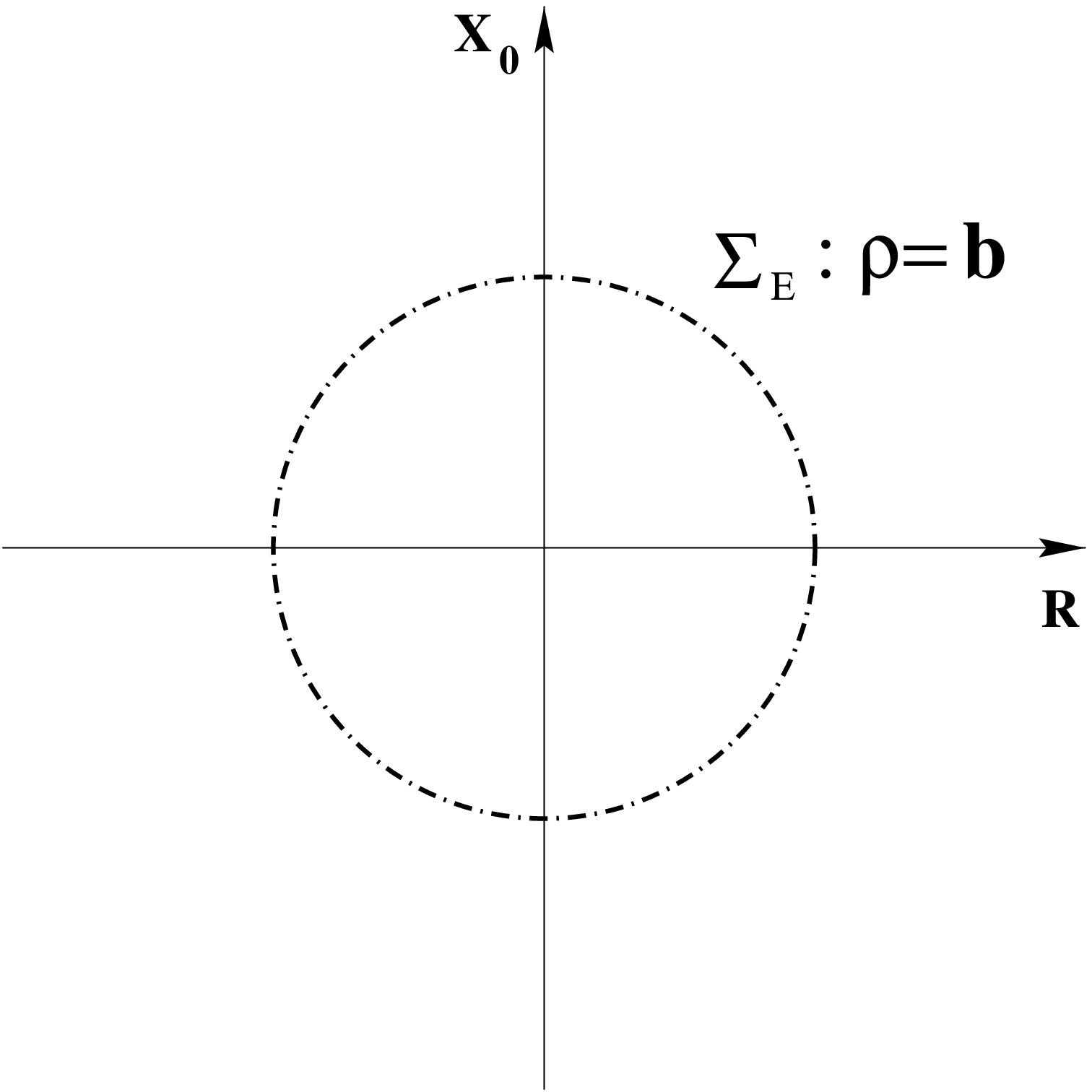, width=6.0cm}}
\hfill
{}
\caption[f1]{The geometry of problem $A$. Solid lines in the left figure
represent motion of spherical semi-transparent mirror $\Sigma$ 
in the spacetime. A dash-dotted line in the right figure corresponds to
the surface $\Sigma_E$ in the Euclidean space obtained by the Wick's
rotation $T\rightarrow iX_0$.}
\end{figure}

As a first example, we consider the case when there exists a single
mirror and its  surface worldsheet $\Sigma$ is described by the
equation
\be
\Sigma \,: \quad R^2 - T^2 = b^2\,,   \n{2.17}
\ee
where
\be
R^2  = \sum_{i=1}^dX_i^2\,.   \n{2.18}
\ee
Such a mirror is spherical. Its radius $R$ changes with time $T$ in
such a  way that points of the mirror are moving with a constant
acceleration  $a=b^{-1}$ orthogonal to the surface of the mirror. In
$(T,R)$-coordinates the motion is represented by a  hyperboloid (see
Fig.1). The Wick's rotation $T\rightarrow iX_0$  transforms $\Sigma$
into $\Sigma_E$
\be
\Sigma_E \,: \quad R^2 + X^2 = b^2\,,   \n{2.19}
\ee
which is a $d$-dimensional sphere $S^d$ of radius $b$ embedded 
into $D$-dimensional Euclidean space.

\subsubsection{Model B}

\begin{figure}[t]
\hfill
{\epsfig{file=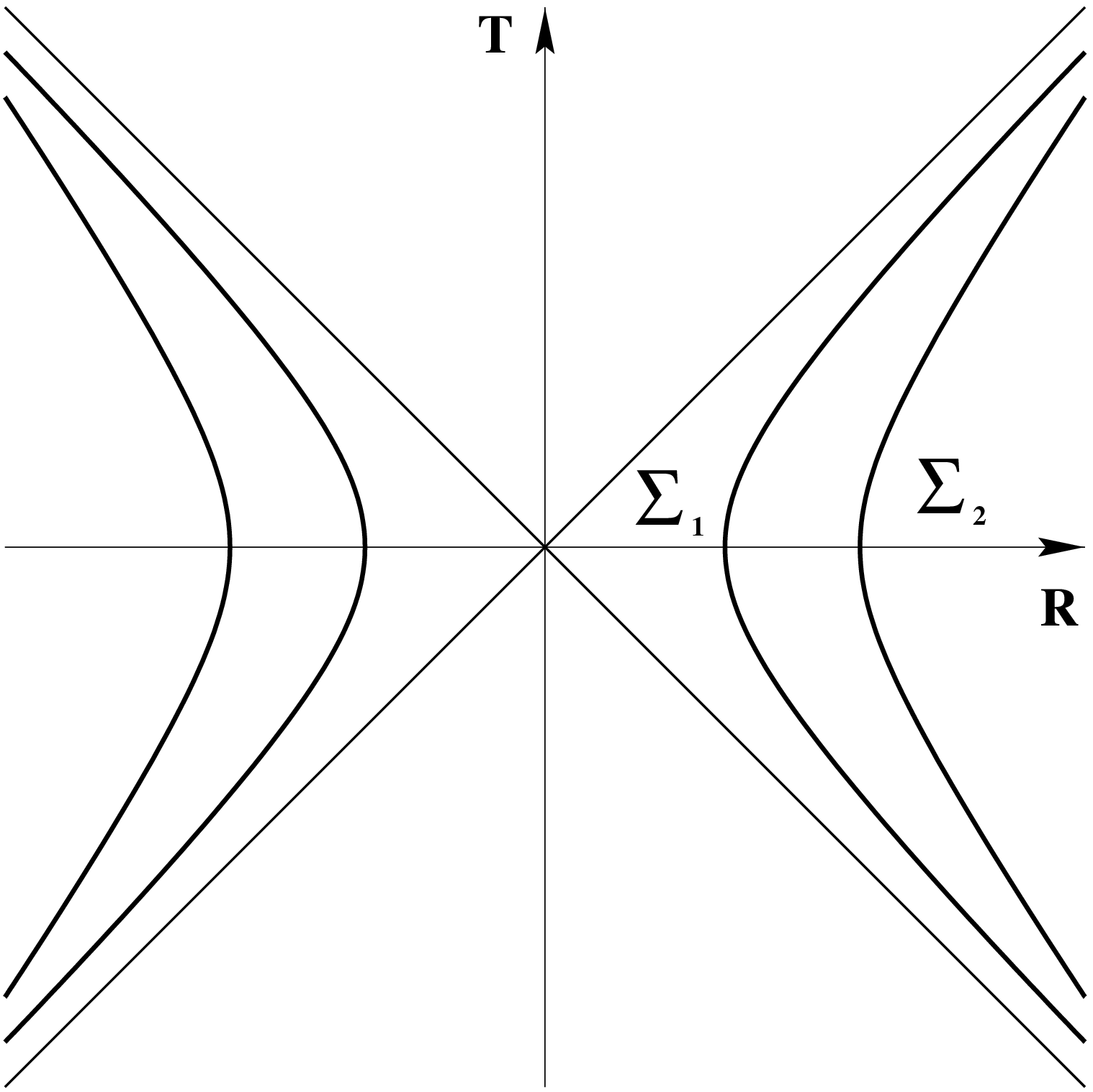, width=6.0cm}}
\hfill 
{\epsfig{file=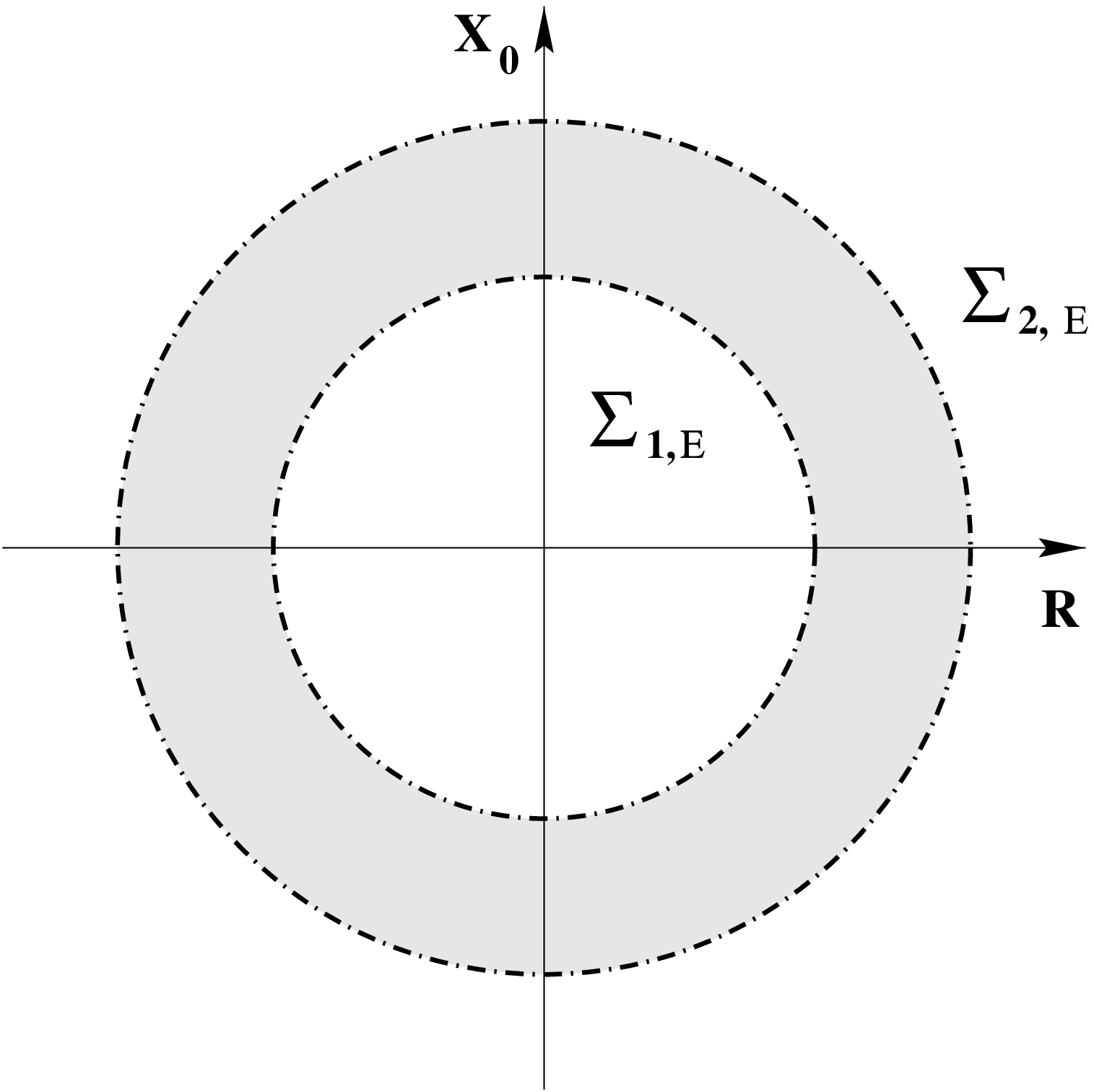, width=6.0cm}}
\hfill
{}
\caption[f2]{The geometry of problem $B$. Solid lines in the left figure
represent motion of spherical semi-transparent mirrors $\Sigma_1$ and
$\Sigma_2$ 
in the spacetime. Dash-dotted lines in the right figure correspond to
the surfaces $\Sigma_{1,E}$ and $\Sigma_{2,E}$ 
in the Euclidean space obtained by the Wick's
rotation $T\rightarrow iX_0$.}
\end{figure} 

Another example is a case when there exist a two concentric
spherical mirrors. In this model, the surface $\Sigma$ consists of two
concentric mirror  surfaces, $\Sigma_1$ and $\Sigma_2,$ given by (see
Fig.2)
\be
\Sigma_1 \,: \quad R^2 - T^2 = b_1^2\,,   \n{2.20}
\ee
\be
\Sigma_2 \,: \quad R^2 - T^2 = b_2^2\,.   \n{2.21}
\ee
We assume that $b_2>b_1$ and call $\Sigma_2$ and $\Sigma_1$ an 
external and  an internal mirror,  respectively. After the Wick's rotation 
$T\rightarrow iX_0,$ these surfaces become (see Fig.2)
\be
\Sigma_{1,E} \,: \quad R^2 + X^2 = b_1^2\,,   \n{2.22}
\ee
\be
\Sigma_{2,E} \,: \quad R^2 + X^2 = b_2^2\,.   \n{2.23}
\ee

\section{Calculation of Green Functions}\label{s3}

\setcounter{equation}0
\subsection{Green Functions in Spherical Coordinates}

We start with the Euclidean formulation. It is convenient to introduce 
spherical coordinates in the $D$-dimensional\footnote{
We assume that $D>2$. A special case of a two-dimensional spacetime
which requires slight modifications will be discussed in Section~5.3.3.
} Euclidean space (\ref{2.9}). 
For this purpose we first parameterize the line element $d\Omega_d^2$ on 
a unit d-dimensional sphere as follows:
\be
d\Omega_1^2 = d\vartheta_1^2 \,,   \n{3.1}
\ee
\be
d\Omega_n^2 = d\vartheta_n^2 + \sin^2\vartheta_n\,d\Omega_{n-1}^2 \,,  
\quad (n=2,\ldots,d)\,.\n{3.2}
\ee
In these co-ordinates
\be
d\Omega_d^2 =  \sum_{i,j=1}^d\Omega_{ij}^{(d)}\,d\vartheta_i\, 
d\vartheta_j\,,   \n{3.3}
\ee

\be
\Omega_{ij}^{(d)} = \mbox{diag}\,(1,\sin^2\vartheta_d(1,\sin^2\vartheta_
{d-1}(\ldots\,.)))\,,   \n{3.4}
\ee
\be
\sqrt{\Omega_d} \equiv \sqrt{\mbox{det}\left(\Omega_{ij}^{(d)}\right)} =
\sin^{d-1}\vartheta_d\, \sin^{d-2}\vartheta_{d-1}\ldots\sin \vartheta_2\,.
\n{3.5}
\ee
We can write the line element (\ref{2.9}) as 
\be
ds_E^2 = d\rho^2 + \rho^2\,d\Omega_d^2\,,   \n{3.6}
\ee
where 
\be
R = \rho\sin\vartheta_d\,, \quad X_0 = \rho\cos\vartheta_d\,.   \n{3.7}
\ee
In what follows, we shall also use the notation
\be
\eta = \vartheta_d\,.   \n{3.8}
\ee
In spherical coordinates, the Euclidean ``Box"-operator $\Box_E$ takes 
the form
\be
\Box_E = \frac{1}{\rho^d}\,\frac{\partial\ }{\partial\rho}\left(\rho^d\frac{
\partial\ }{\partial\rho}\right) + \frac{1}{\rho^2}\,\Delta_d\,,   \n{3.9}
\ee
where $\Delta_d$ is the Laplace-Beltrami operator on $S^d$.

Scalar spherical harmonics on $S^d$, which are solutions of the eigenvalue 
problem
\be
\Delta_d\,Y = \lambda Y   \n{3.10}
\ee
on $S^d$, have been studied by many authors 
(see, e.g. \cite{ChMy:84, Ratr:85, Higu:87}). 
Let $l_1, l_2, \ldots \,, l_d$ be integers that satisfy
\be
l_d\ge l_{d-1}\ge \ldots \,\ge l_2\ge |l_1|\,,  \n{3.11}
\ee
and denote
\[
{}_n\hat{P}_{\lambda}^l(\vartheta ) = \left[\frac{2\lambda + n -1}{2}\,\frac
{(\lambda +l+n-2)!}{(\lambda -l)!}\right]^{1/2}(\sin\vartheta )^{-(n-2)/2}
\]
\be
\hspace{-2.4cm}
\cdot P_{\lambda +(n-2)/2}^{-(l+(n-2)/2)}(\cos\vartheta )\,,  \n{3.12}
\ee
where $P_{\nu}^{\mu}(x)$ is the associated Legendre function of the first kind. 
Then the spherical harmonics on $S^d$ are given by \cite{Higu:87}
\be
Y_{l_d\ldots \,l_1}(\vartheta_d,\ldots ,\vartheta_1) = \left[\prod _{n=2}^
d {}_n\hat{P}_{l_n}^{l_{d-1}}(\vartheta_n )\right]\frac{e^{il_1\vartheta_1}}
{\sqrt{2\pi}}\,.  \n{3.13}
\ee
Scalar spherical harmonics are normalized as
\be
\int d\vartheta_1\ldots d\vartheta_d\,\sqrt{\Omega_d}\,Y_{l_d\ldots \,l_1}
\cdot \bar{Y}_{l'_d\ldots \,l'_1} = \delta_{l_d\,l'_d}\cdots \delta_{l_1l'_1}
\,,  \n{3.14}
\ee
and form a complete set of functions on $S^d$. Scalar spherical harmonics 
obey the relation
\be
\Delta_d\,Y_{l_d\ldots \,l_1} = -l_d\,(l_d+d-1)
Y_{l_d\ldots \,l_1} \,.  \n{3.15}
\ee

To make the notations brief, we denote
\be
L =l_d\,, \quad W = \{l_{d-1},\ldots ,l_1\} \,,  \n{3.16}
\ee
where values of $l_n,$ which enter the collective index $W,$ are restricted 
by (\ref{3.11}).
In these notations equations (\ref{3.15}) and  (\ref{3.14}) take the form
\be
\Delta_d\,Y_{LW} = -L(L+d-1)Y_{LW} \,, \hspace{0.5cm} (L\ge 0), \n{3.17}
\ee
\be
\int d\Omega_d\,Y_{LW}(\Omega )\, \bar{Y}_{L'W'}(\Omega ') = 
\delta_{LL'}\,\delta_{WW'} \,.  \n{3.18}
\ee
Here 
\be
d\Omega_d = d\vartheta_1\ldots d\vartheta_d\sqrt{\Omega_d} \,,  \n{3.19}
\ee
and
\be
\delta_{WW'} = \delta_{l_{d-1}l_{d-1}'}\ldots \delta_{l_1l_1'} \,.  \n{3.20}
\ee
 We also have the relation
\be
\sum_{L=0}^{\infty}\sum_WY_{LW}(\Omega )\, \bar{Y}_{LW}
(\Omega ') = \delta(\Omega ,\Omega') \,,  \n{3.21}
\ee
 which is a consequence of the 
completeness of scalar spherical harmonics, and where 
$\delta(\Omega ,\Omega')$ is an invariant $\delta$-function on unit 
sphere $S^d$.

The external potential $V_{\Sigma_E}$ for both of the problems, $A$ and 
$B$, is evidently invariant under rotations in the Euclidean space. That 
is why both  problems allow for separation of variables. Substituting 
(\ref{3.21}) into the equation
\be
(\Box_E - V_{\Sigma_E})G_E(x,x') = -\delta(x ,x') \,,  \n{3.22}
\ee
we can obtain the following representation for the Euclidean Green
function:
\be
G_E(x,x') = \sum_{L=0}^{\infty}\sum_W{\cal G}_L(\rho ,\rho')\,Y_{LW}
(\Omega )\,Y_{LW}(\Omega' ) \,.  \n{3.23}
\ee
The radial Green function ${\cal G}_L(\rho ,\rho')$ obeys the equation
\be
\left[\frac{d\ }{d\rho}\left(\rho^d\,\frac{d\ }{d\rho}\right) - \rho^{d-2}\,L(L+d-1) 
-\rho^d\,V_{\Sigma}(\rho )\right]{\cal G}_L(\rho ,\rho') = -\delta (\rho -\rho') \,. 
 \n{3.24}
\ee
Equation (\ref{3.24}) is self-adjoint, and hence ${\cal G}_L$ is a symmetric 
function of its arguments
\be
{\cal G}_L(\rho ,\rho') = {\cal G}_L (\rho', \rho) \,.  \n{3.25}
\ee
For problem $A$
\be
V_{\Sigma}(\rho ) = V_0\,\delta(\rho - b) \,,  \n{3.26}
\ee
while for problem $B$
\be
V_{\Sigma}(\rho ) = V_1\,\delta(\rho - b_1) + V_2\,\delta(\rho - b_2)
\,.  \n{3.27}
\ee

The radial Green function ${\cal G}_L$ does not depend on the collective
index $W$. One can use the following relation to make the summation over
$W$ in (\ref{3.23}) (\cite{Ratr:85}, equation (B.12)):
\be
\sum_WY_{LW}(\Omega )\,\bar{Y}_{LW}(\Omega ') = 
\eta_d\,(2L+d-1)\,
C_L^{(d-1)/2}(\cos \gamma ) \,, \n{3.28}
\ee
where $\gamma $ is the  angle between $\Omega$ and $\Omega'$ on the
unit sphere $S^d$, and 
\be\n{3.28a}
\eta_d={\displaystyle{\frac{\Gamma \left(\frac{d-1}{2}\right)}{4\pi^{(d+1)/2}}
}}\, .
\ee
Note that the volume of the $d$-dimensional unit sphere is
\be
{\cal V}_d={\displaystyle{{2\pi^{{d+1\over 2}}\over 
\Gamma\left( {d+1\over 2}\right)}
}}\, ,
\ee
and hence
\be
\eta_d ={1\over {(d-1)\, {\cal V}_d} }\, .
\ee

The function  $C_L^{(d-1)/2}(x)$ is the Gegenbauer polynomial
 related to the hypergeometric function $F(a,b;c;z)$ as follows
\be\n{3.29}
C_L^{(d-1)/2}(x)={(L+d-2)!\over L! (d-2)!}\, F(-L,L+d-1;{d\over
2};{1-x\over 2})\, .
\ee
Since the hypergeometric function is normalized so
that $F(a,b;c;z=0)=1$, we get
\be
C_L^{(d-1)/2}(1) =  \frac{(L+d-2)!}{L!\,(d-2)!} \,. \n{4.2}
\ee
Thus, we have as the following representation for $G_E$:
\be
G_E(x,x') = 
\eta_d\,
\sum_{L=0}^{\infty} (2L+d-1){\cal G}_L(\rho ,\rho')\,
C_L^{(d-1)/2}(\cos \gamma ) \,.  \n{3.30}
\ee
In this expression, $\gamma$ is a geodesic distance between two points
$(\vartheta_1, \, \ldots\, \vartheta_d)$ and  
$(\vartheta_1^{'}, \, \ldots\, \vartheta_d^{'})$ on the unit sphere. It is
defined as $\gamma=\gamma_d$ by the following relations
\be
\gamma_1 =\vartheta_1-\vartheta_1^{'} \, ,
\ee
\be
\cos \gamma_n =\cos \vartheta_n\, \cos \vartheta_n^{'} 
+\sin \vartheta_n\, \sin \vartheta_n^{'} \, \cos \gamma_{n-1}\,
,\hspace{0.5cm} n=2, \ldots , d\, .
\ee
An ambiguity in the choice of the spherical coordinates connected with
rigid rotations of space can be used to put $\vartheta_1 =\ldots\, =
\vartheta_{d-1} =0$ and $\vartheta_1^{'} = \ldots\, =
\vartheta_{d-1}^{'}=0)$ for any chosen pair of points on the unit
sphere.

\subsection{Problem $A$}

In order to construct ${\cal G}_L (\rho, \rho')$ for problem $A$ , we
shall use  solutions of the homogeneous equation
\be
\left[\frac{d\ }{d\rho}\left(\rho^d\,\frac{d\ }{d\rho}\right) - \rho^{d-2}\,L(L+d-1) 
\right]R_L = 0 \,.  \n{3.31}
\ee
It is easy to verify that this equation has the two linearly 
independent solutions, $\rho^{L_+}$ and  $\rho^{L_-}$, where
\be
L_+ = L\,, \quad L_- = -(L+d-1)\,.  \n{3.32}
\ee
The solution $\rho^{L_+}$ is regular at the origin $ \rho = 0$, while
the solution   $\rho^{L_-}$ decreases to zero at infinity. The Green
function   ${\cal G}_L(\rho ,\rho')$ which is regular at $\rho = 0$ and
decreases to zero at  infinity is of the form 
\be
{\cal G}_L(\rho ,\rho') = \left\{
\begin{array}{ll}
(A_-\,\rho^{L_-} + A_+\,\rho^{L_+})(\rho ')^{L_+}\,, & 
\quad \ b>\rho >\rho '\ge 0\,, \\
\  & \\
B\rho^{L_-} \,(\rho ')^{L_+}\,, & 
\quad \rho >b >\rho '\ge 0\,, \\
\  & \\
\rho^{L_-}\left(C_-\,(\rho')^{L_-} + C_+\,(\rho ')^{L_+}\right), & 
\quad \rho >\rho' > b \ge 0\,.
 \end{array} \right.  \n{3.33}
\ee
The symmetry condition (\ref{3.25}) determines ${\cal G}_L(\rho ,\rho')$ for 
other possible relative positions of $\rho$ and $\rho'$.

The radial Green function ${\cal G}_L(\rho ,\rho')$ is continuous, while its 
first derivatives  have jumps. The form and location of the jumps directly 
follow from equation (\ref{3.24}) with potential (\ref{3.26}). The jump at 
$\rho = \rho'$ is of the form
\be
\left[\frac{d{\cal G}_L(\rho ,\rho')}{d\rho}\right]_{\rho = \rho'+0} - 
\left[\frac{d{\cal G}_L(\rho ,\rho')}{d\rho}\right]_{\rho = \rho'-0} = 
-\,\frac{1}{(\rho')^d} \,.  \n{3.34}
\ee
For $\rho' \neq b,$ the jump at $\rho = b$ is
\be
\left[\frac{d{\cal G}_L(\rho ,\rho')}{d\rho}\right]_{\rho = b+0} - 
\left[\frac{d{\cal G}_L(\rho ,\rho')}{d\rho}\right]_{\rho = b-0} = 
V_0\,{\cal G}_L(b,\rho') \,.  \n{3.35}
\ee
The jump at $\rho' = b$ and $\rho \neq b$ has a similar form. Relation
(\ref{3.34}) allows us to find coefficients $A_-$ and  $C_+$ in
(\ref{3.33}). Continuity of ${\cal G}_L(\rho ,\rho')$ at $\rho = b$
and  $\rho' = b,$ together with relation  (\ref{3.35}), fix the other
coefficients.  By solving the corresponding linear equations, we get
\be
\begin{array}{l}
A_- = C_+ ={\displaystyle \frac{1}{2(L+\beta_d)}}\,,  \hspace{0.5cm}
B =  {\displaystyle\frac{1}{2(L+\beta_d+U_0)}} \,, \\
\    \\ 
A_+ = {\displaystyle -\,\frac{U_0}{2(L+\beta_d)(L+\beta_d+U_0)}{1\over
b^{2L+d-1}}}\,, \\
\   \\
C_- = {\displaystyle -\,\frac{U_0}{2(L+\beta_d)(L+\beta_d+U_0)}}
b^{2L+d-1}\, .
 \end{array}   \n{3.36}
\ee
Here we introduced notations
\be \n{3.37}
\beta_d={d-1\over 2}\, ,\hspace{0.5cm} U_0={1\over 2}bV_0\, .
\ee

In the absence of a mirror, when $V_0 = 0$, relations (\ref{3.33}) are simplified
to become
\be
A_- = C_+ = B = \frac{1}{2L+d-1}\,,  \quad A_+ = C_- = 0\,.  \n{3.38}
\ee
In this limit, the Green function $G_E(x,x')$ coincides with the free Green 
function (\ref{2.11}). This can be easily verified by using the relation 
(see e.g. \cite{Ratr:85}, eq. (B.13))
\be
\frac{1}{|X-X'|^{d-1}} = {1\over \eta_d}
\sum_{L=0}^{\infty}\sum_W \frac{\rho_>^{L_-}\,\rho_<^{L_+}}{2L+d-1}\,
Y_{LW}(\Omega )\,\bar{Y}_{LW}(\Omega')\,.  \n{3.39}
\ee
Here $\rho_> = \mbox{max}\,(\rho ,\rho')$ and 
$\rho_< = \mbox{min}\,(\rho ,\rho')$.

Subtracting the free Green function $G_0(x,x')$ from $G_E(x,x')$ gives the 
renormalized Green function
\[
G_E^{\ind{ren}}(x,x') = \sum_{L=0}^{\infty}
\sum_W{\cal G}_L^{\ind{ren}}(\rho ,\rho')\,
Y_{LW}(\Omega )\,\bar{Y}_{LW}(\Omega')\,,  
\]
\be\n{3.40}
=\eta_d
\sum_L (2L+d-1)\, {\cal G}_L^{\ind{ren}}(\rho ,\rho')\, C^{(d-1)/2}_{L} 
(\cos(\gamma))\, .
\ee
Here
\be
{\cal G}_L^{\ind{ren}}(\rho ,\rho') = -\,\frac{U_0}{2(L+\beta_d)(L+\beta_b+U_0)}
\,\frac{1}{b^{d-1}}\left\{ 
\begin{array}{ll}
\displaystyle \left(\frac{\rho \rho '}{b^2}\right)^{L_+}, & 
\ b>\rho  ,\ \rho '\ge 0\,, \\
\  & \\
\displaystyle \left(\frac{\rho \rho '}{b^2}\right)^{L_-}, & 
\ \rho ,\ \rho '\ge b\,. 
 \end{array} \right.   \n{3.41}
\ee

\subsection{Problem $B$}

The calculation of the Green function for the problem $B$ is similar to
the  calculations of the previous subsection. We again use the
representation  (\ref{3.23}) and write the radial Green function ${\cal
G}_L(\rho ,\rho')$ in the  form\footnote{ Coefficients $A_{\pm},
B_{\pm}, \ldots$ which enter relations (\ref{3.42}) -- (\ref{3.47}),
(\ref{3.51}) and (\ref{3.52}), depend on the multipole moment $L$. In
order to simplify formulas we do not indicate this dependence
explicitly.}
\be
{\cal G}_L(\rho ,\rho') = \left\{ 
\begin{array}{ll}
\left(A_-\,\rho^{L_-} + A_+ \,\rho ^{L_+}\right)(\rho')^{L_+}, & 
\ b_1>\rho >\rho '\ge 0\,, \\
\  & \\
\left(B_-\,\rho^{L_-} + B_+ \,\rho ^{L_+}\right)(\rho')^{L_+}, & 
\ b_2>\rho >b_1>\rho '\ge 0, \\
\  & \\
C\,\rho^{L_-}(\rho')^{L_+}\,, & 
\ \rho >b_2>b_1>\rho '\ge 0\,, \\
\  & \\
C^{-1}\, \left(D_-\,\rho^{L_-} + D_+ 
\,\rho ^{L_+}\right)\!\left(E_-(\rho')^{L_-}+E_+(\rho')^
{L_+}\right)\!, & 
\ b_2>\rho >\rho '>b_1> 0\,, \\
\  & \\
\rho^{L_-}\left(F_-(\rho')^{L_-}+F_+(\rho')^{L_+}\right), & 
\ \rho >b_2>\rho '>b_1> 0\,, \\
\  & \\
\rho^{L_-}\left(G_-(\rho')^{L_-}+G_+(\rho')^{L_+}\right), & 
\ \rho >\rho '>b_2>b_1> 0 \,. 
 \end{array} \right.  \n{3.42}
\ee
The expression for ${\cal G}_L(\rho,\rho'$ for other possible positions
of $\rho$ and  $\rho'$ follows from the symmetry of this function. As
before, jump conditions  (\ref{3.35}) and (\ref{3.36}) allow us to
obtain the unknown coefficients  which enter relation (\ref{3.42}).
After straightforward but long calculations  we find that
\be\n{3.43}
A_+  =
-{1\over 2\Omega_L\, b_1^{2L+d-1}} \left[ U_1 + U_2\, \gamma_L
 + {U_1\, U_2 \over L+\beta_d} \, 
\left(1-\gamma_L \right)
\right]\, , 
\ee
\be\n{3.44}
A_-  = G_+ = \frac{1}{2(L + \beta_d)}\, ,\qquad
C =   \frac{L+\beta_d}{2\Omega_L} \, ,
\ee
\be\n{3.45}
B_+ = D_+=- \frac{U_2}{2\Omega_L}\frac{1}{b_2^{2L + d - 1}}\, ,
\qquad
B_- = D_- = \frac{L + \beta_d + U_2}{2\Omega_L}\, ,
\ee
\be\n{3.46}
E_+ = F_+ =   \frac{L + \beta_d + U_1}{2\Omega_L}\, , \qquad
E_- = F_-  =  -\frac{ U_1}{2\Omega_L}{b_1^{2L + d-1}}\, ,
\ee
\be\n{3.47}
G_-=
-{b_2^{2L+d-1}\over 2\Omega_L} \left[ U_1\, \gamma_L + U_2\,
 + {U_1\, U_2 \over L+\beta_d} \, 
\left(1-\gamma_L \right)
\right]\, .
\ee
Here, we used notations
\be\n{3.48}
U_1 = \frac{b_1 V_1}{2}, \qquad U_2 \ = \ \frac{b_2 V_2}{2}\, ,
\qquad \gamma_L =\left( {b_1\over b_2} \right)^{2L+d-1}\, ,
\qquad \beta_d ={d-1 \over 2}\, ,
\ee 
\be\n{3.49}
\Omega_L  =  (L + \beta_d + U_1)(L + \beta_d + U_2) - U_1 U_2
\gamma_L \, .
\ee

For calculation of $\langle \hat{\varphi}^2\rangle^{\ind{ren}}$ and
$\langle \hat{T}^{\nu}_{\mu}\rangle^{\ind{ren}}$ we need ${\cal
G}_L^{\ind{ren}},$ which can be obtained from (\ref{3.42}) by subtracting
the vacuum radial Green function ${\cal G}_L^{0}$. The latter can be
obtained by setting
$U_1=U_2=0$ in (\ref{3.42}). In the regions of interest ($\rho'<\rho<b_1$ (inner),
$\rho>\rho'>b_2$ (outer), and $ b_2>\rho>\rho'>b_1$ (intermediate)) the
vacuum radial Green function has the same form
\be
{\cal G}_L^{0}(\rho,\rho')={1\over 2(L+\beta_d)}\, \rho^{L_-}\,
{\rho'}^{L_+}\, .
\ee
As a result, we get the following expressions for ${\cal
G}_L^{\ind{ren}}$ :\\

\noindent
{\bf Inner region}  $\rho, \rho^\prime <b_1$:
\be\n{3.50}
{\cal G}_L^{\ind{ren}} ( \rho, \rho')  = A_+ (\rho \rho')^L\, ;
\ee

\medskip

\noindent
{\bf Outer region} $ \rho ,\rho^\prime > b_2 $:
\be\n{3.51}
{\cal G}_L^{\ind{ren}} ( \rho, \rho')  = 
{G_-\over (\rho \rho')^{L+d-1}}\, ;
\ee

\medskip

\noindent
{\bf Intermediate region}  $ b_2 > \rho , \rho^\prime >  b_1$:
\[
{\cal G}_L^{\ind{ren}} ( \rho, \rho')  = 
-{1 \over 2\Omega_L\, (L+\beta_d)}
\left[
U_1\, (L+\beta_d +U_2)\, {b_1^{2L+d-1}\over (\rho\rho')^{L+d-1}}
\right.
\]
\be\n{3.52}
\left.
\qquad
+
U_2\, (L+\beta_d +U_1)\, { (\rho\rho')^{L}\over b_2^{2L+d-1}}
-
U_1\, U_2\, \gamma_L\, \left( \frac{\rho^L}{(\rho^\prime)^{L + d - 1}} + 
 \frac{(\rho^\prime)^L}{\rho^{L + d - 1}} \right)\, 
\right] \, .
\ee

\bigskip

As expected, ${\cal G}_L^{\ind{ren}} ( \rho, \rho')$ is a symmetric
function of its arguments.
It is easy to verify that when one of the potentials, $U_1$ or $U_2$,
vanishes,  expressions (\ref{3.50})--(\ref{3.52}) 
for ${\cal G}_L^{\ind{ren}}$ reduce to
(\ref{3.41}). In the other limit of perfectly reflecting mirrors,
when $U_1=U_2=\infty$, relations (\ref{3.50})--(\ref{3.52}) correctly
reproduce the result of \cite{FrSe:79}.

\section{Calculation of $\langle\hat{\varphi}^2\rangle^{\ind{ren}}$}\label{s4}
\setcounter{equation}0

\subsection{Problem $A$}

\subsubsection {Expression for
$\langle\hat{\varphi}^2\rangle^{\ind{ren}}$}

In order to calculate $\langle\hat{\varphi}^2\rangle^{\ind{ren}}$, we need to 
obtain $G_{\!E}^{\ind{ren}}(x,x')$ in the coincidence limit $x'\rightarrow x$
\be
\langle\hat{\varphi}^2\rangle^{\ind{ren}} = G_{\!E}^{\ind{ren}}(x,x)\,. \n{4.1}
\ee
Let us discuss problem $A$ at first. We shall use representation
(\ref{3.30}) for the Green function, where, instead of ${\cal G}_L,$ we
substitute its renormalized version ${\cal G}_L^{\ind{ren}}$ given by
(\ref{3.41}).  In order to find a coincidence limit, we must put
$\gamma =0$ and $\rho=\rho'$.  Using relations  (\ref{4.2}), (\ref{3.30}), 
and (\ref{3.41})  we get 
\be\n{4.3}
\langle\hat{\varphi}^2(\rho )\rangle^{\ind{ren}} = 
-\,\frac{U_0\, \eta_d}{b^{d-1}}\, 
\left\{ 
\begin{array}{ll}
F^{(d)}((\rho/b)^2, U_0+(d-1)/2)\, , & 
\ \rho < b\,, \\
\  & \\
\left({b\over \rho}\right)^{2(d-1)} 
F^{(d)}((b/\rho)^2, U_0+(d-1)/2)\, . & 
\ b>\rho ; \\
\end{array}
\right.
\ee
where the function $F^{(d)}$ is 
\be
F^{(d)}(z,\beta )  = \sum_{L=0}^{\infty} \frac{(L+d-2)!}{L!\,(d-2)!\,(L+\beta )}
\,z^{L} \,. \n{4.6}
\ee
Some properties of this function  are discussed in Appendix~A.
In particular, relation (\ref{A.3}) allows one to show that 
at the center $\rho = 0$
\be
\langle\hat{\varphi}^2(\rho = 0)\rangle^{\ind{ren}} =
 -\, \frac{\eta_d}{b^{d-1}}\,
\frac{U_0}{ (d-1)/2+U_0}\,. 
\n{4.9}
\ee

Quantity $\langle\hat{\varphi}^2(\rho )\rangle^{\ind{ren}}$ is
divergent at $\rho=b$. Using (\ref{A.5}) one gets the following
expression for the leading divergent part of 
$\langle\hat{\varphi}^2(\rho )\rangle^{\ind{ren}}$ at $\rho=b$
\be\n{4.12}
\langle\hat{\varphi}^2(\rho)\rangle^{\ind{ren}}\sim
-{U_0\over b^{d-1}}{\eta_d\over (d-2)}{1\over (1-x^2)^{d-2}}\, ,
\ee
where
\be\n{4.13}
x= \left\{ 
\begin{array}{ll}
\rho/b, & 
\ \rho\le b\,, \\
\  & \\
b/\rho, & 
\ b>\rho . \\
\end{array}
\right.
\ee

In the limiting case $U_0 = \infty$,
$\langle\hat{\varphi}^2(x) \rangle^{\ind {ren}} $ can be expressed in
terms of elementary functions. Using (\ref{A.6}) we can rewrite
(\ref{4.3}) for $U_0 = \infty$ in the form
\be\n{4.15}
\langle\hat{\varphi}^2(x)\rangle^{\ind{ren}} = -\, 
\frac{\eta_d}{b^{d-1}}\,\frac{1}{|1-\left(\rho /b\right)^2|^
{d-1} }\,. 
\ee
For $d=3,$ this result directly follows from expression (\ref{3.5}) of
Ref. \cite {FrSe:79} for the Green function of a scalar massless field
in the presence  of an ideal mirror with the same choice of the surface
$\Sigma$ as for our  problem $A$. For $d\neq 3$ the result 
(\ref{4.15}) can be easily verified by  the method of images which was
used in Ref. \cite{FrSe:79}.

\subsubsection{$4$-dimensional case}

\begin{figure}
\centerline{\epsfig{file=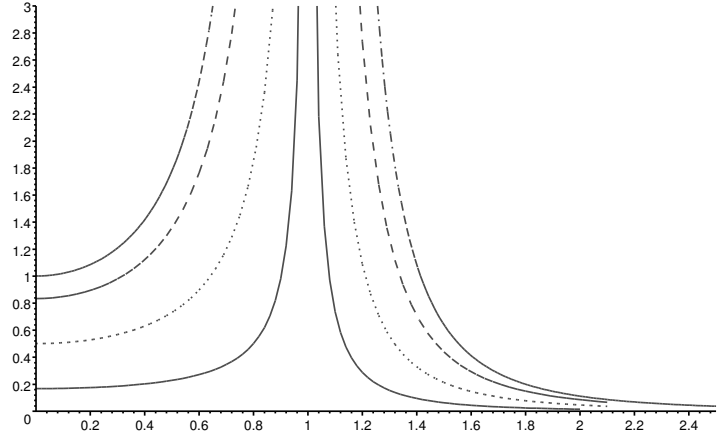, width=10cm}}
\caption[f3]
{$-4\pi^2 b^2 \langle \hat{\varphi}^2\rangle^{\ind{ren}}$ in
4-dimensional spacetime as the function of $\rho/b$ for different values
of the parameter $U_0$: solid line -- $U_0=0.2$, dotted line --
$U_0=1.0$, and dashed line -- $U_0=5.0$. The dashed and dotted line
corresponds to an ideally reflecting mirror ($U_0=\infty$).}
\label{f3}
\end{figure} 
\begin{figure}
\centerline{\epsfig{file=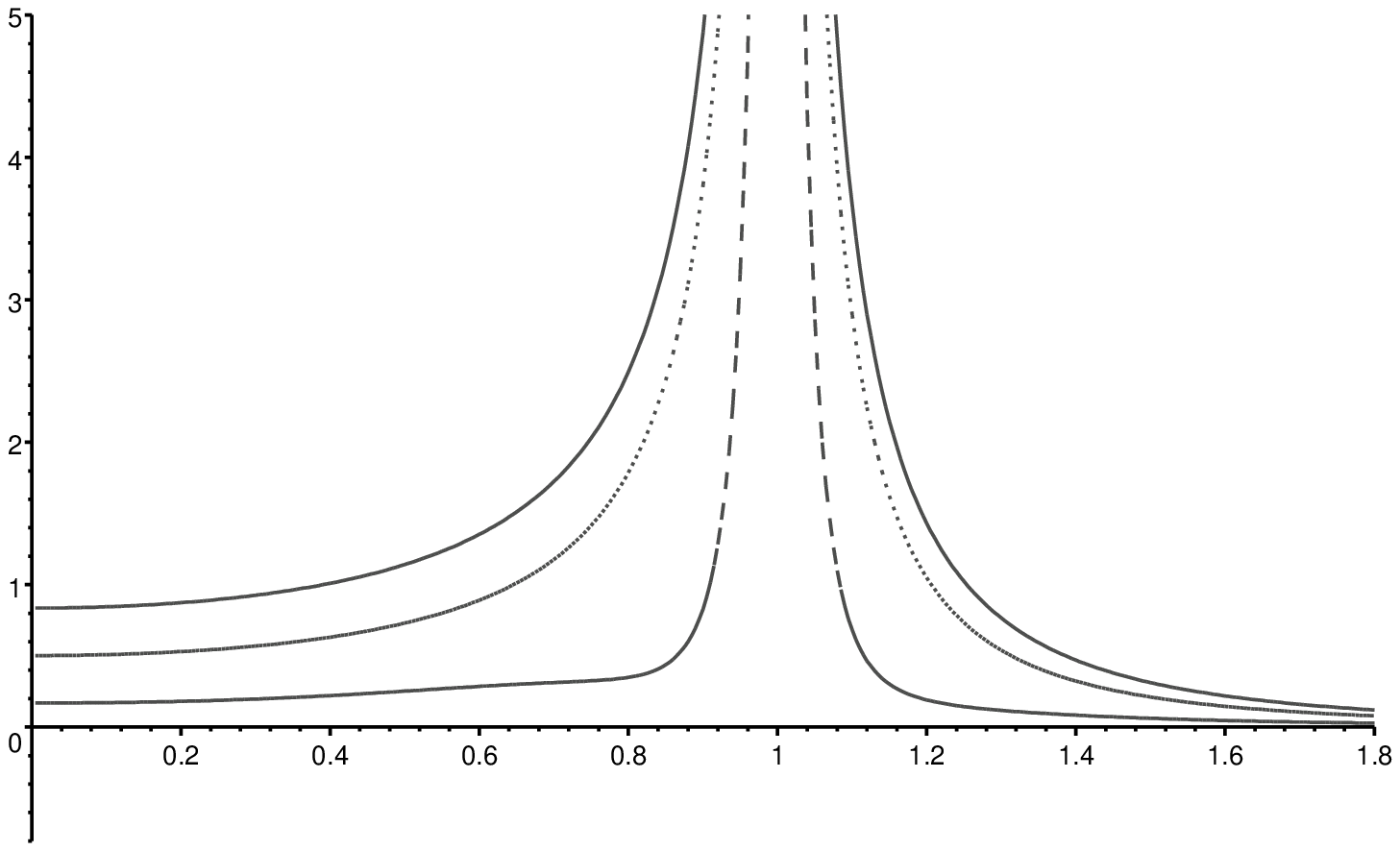, width=10cm}}
\caption[f4]
{Function $H(\rho,U_0)$  for different values
of the parameter $U_0$: solid line -- $U_0=0.2$, dotted line --
$U_0=1.0$, and dashed line -- $U_0=5.0$. }
\label{f4}
\end{figure} 
\begin{figure}
\centerline{\epsfig{file=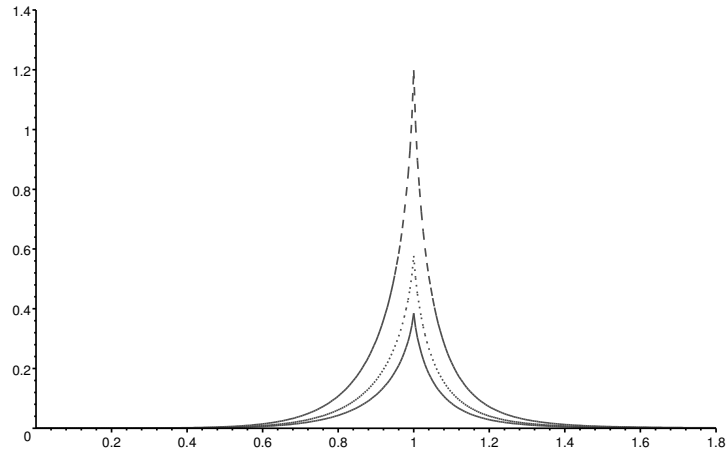, width=10cm}}
\caption[f5]
{Function $G(\rho,U_0)$  for different values
of the parameter $U_0$: solid line -- $U_0=0.2$, dotted line --
$U_0=1.0$, and dashed line -- $U_0=5.0$. }
\label{f5}
\end{figure}

Now we discuss  some properties of
$\langle\hat{\varphi}^2\rangle^{\ind{ren}}$ in the case of the
4-dimensional spacetime, which has the most physical interest. Figure
3 shows plots of $\langle\hat{\varphi}^2\rangle^{\ind{ren}}$ (for
$d=3$)  as a  function of $\rho /b$  for different  values of the
height of the potential $U_0$. For comparison, it also contains 
$\langle\hat{\varphi}^2\rangle^{\ind{ren}}$ for an ideally reflecting
mirror,  (\ref{4.15}).

It is evident that, near the mirror surface, the dominant divergent part
is directly related to the singularity of the potential. In order to
extract the singular part of $\langle\hat{\varphi}^2\rangle^{\ind{ren}},$
we represent  $F^{(3)}$ in the form
\be\n{4.16}
F^{(3)}(z,\beta)=f(z,\beta)+g(z,\beta)\, ,
\ee
where 
\be\n{4.17}
h(z,\beta)={1\over 1-z} -(1-\beta)\ln (1-z)+{1-\beta \over \beta}
+\left[ {2\over 1+\beta}-(2-\beta)\right]\, z
+\left[ {3\over 2+\beta}+{\beta -3\over 2}\right]\, z^2
\, .
\ee
\[
g(z,\beta)=-(1-\beta)\tilde{g}(z,\beta)\, ,
\]
\be\n{4.18}
\tilde{g}(z,\beta)=
\beta \sum_{L=1}^{\infty} {z^L\over L(L+\beta)}
+{1\over 1-\beta}\left\{ 
\left[ {2\over 1+\beta}-(2-\beta)\right]\, z
+\left[ {3\over 2+\beta}+{\beta -3\over 2}\right]\, z^2
 \right\}
\, .
\ee
The series in the right-hand side of (\ref{4.18}) is convergent at
$z=1$. Hence, the function $f(z,\beta)$ contains all the divergences of
$F^{(3)}$ at $z=1$. The last terms in the right-hand side of (\ref{4.17}) 
which are proportional to $z$ and $z^2$
are finite at $z=1$. They are added in order for $h(z,\beta)$ to have 
the same first 3 terms in the Taylor series expansion at $z=0$ as the
original function $F^{(3)}$. Hence, we can write
$\langle\hat{\varphi}^2\rangle^{\ind{ren}}$ in the form
\be\n{4.19}
-{4\pi^2b^2\over U_0}\langle\hat{\varphi}(\rho)^2\rangle^{\ind{ren}}
= H(\rho,U_0)+U_0 G(\rho,U_0)\, ,
\ee
where
\be\n{4.20}
H(\rho,U_0)=
\left\{ 
\begin{array}{ll}
h((\rho/b)^2,1+U_0), & 
\ \rho\le b\,, \\
\  & \\
(b/\rho)^4 h((b/\rho)^2, 1+U_0), & 
\ b>\rho ; \\
\end{array}
\right.
\ee

\be\n{4.21}
G(\rho,U_0)=
\left\{ 
\begin{array}{ll}
\tilde{g}((\rho/b)^2, 1+U_0), & 
\ \rho\le b\,, \\
\  & \\
(b/\rho)^4 \tilde{g}((b/\rho)^2, 1+U_0), & 
\ b>\rho . \\
\end{array}
\right.
\ee
The plots of functions $H$ and $G$ for different values of $U_0$ are
shown in Figures~\ref{f4} and \ref{f5}.

\subsection{Problem $B$}

Using relations (\ref{4.1}), (\ref{4.2}), and
(\ref{3.50})--(\ref{3.52}),
we can write the following representation for
$\langle\hat{\varphi}^2\rangle^{\ind{ren}}$ for the two-mirror case
(problem B):
\be\n{4.22}
\langle\hat{\varphi}^2\rangle^{\ind{ren}} ={\eta_d\over (d-2)!} 
\sum_{L=0}^{\infty}
{(L+\beta_d)(L+d-2)!\over L!} {\cal R}_L (\rho)\, ,
\ee
where
\be\n{4.23}
{\cal R}_L (\rho)=
\left\{
\begin{array}{ll}
\displaystyle{
-{1\over \Omega_L\, b_1^{d-1}}\, 
\left({\rho\over b_1}\right)^{2L}\,
\left[ 
U_1+U_2\, \gamma_L
+
{U_1\, U_2 \over L+\beta_d}\, 
\left( 1-
\gamma_L
\right)
\right]
}\, , 
& \rho\le b_1\,,
 \\ 
\  & \\
\displaystyle{
-{1\over \Omega_L(L+\beta_d)}{1\over \rho^{d-1}} 
\left[
U_1 (L+\beta_d+U_2) \left( {b_1\over \rho}\right)^{2L+d-1}
\right.
}
 & \\
\  & \\
\left.
\qquad
+
\displaystyle{
U_2 (L+\beta_d+U_1) \left( {\rho\over b_2}\right)^{2L+d-1}-
2U_1 U_2 \gamma_L
}
\right]
\, , &
{}\hspace{-0.5cm}b_1\le \rho\le b_2\, ,
 \\
\  & \\ 
\displaystyle{
-{1\over \Omega_L\, \rho^{d-1}}\left({b_2\over \rho}\right)^{2L+d-1}\,
\left[ 
U_2+U_1\, \gamma_L
+
{U_1\, U_2 \over L+\beta_d}\, 
\left( 1-
\gamma_L
\right)
\right]
},& 
\rho\ge b_2 .
\end{array}
\right.
\ee
and $\gamma_L = (b_1/ b_2)^{2L+d-1}$.

Notice that only the $L=0$ modes contribute to 
$\langle \hat{\varphi}^2\rangle^{\ind{ren}}$ at the origin. Thus, one has
\be\n{4.25}
\langle\hat{\varphi}^2(\rho = 0)\rangle^{\ind{ren}} =
 -\, \frac{\eta_d}{b_1^{d-1}}\,
{\beta_d (U_1+ \gamma_0)+ U_1 U_2 (1-\gamma_0)
\over
(\beta_d +U_1)(\beta_d+U_2)-U_1U_2 \gamma_0}\, ,
\ee
where
$
\gamma_0 = (b_1/b_2)^{d-1}\, 
$ .

\begin{figure}
\begin{tabular}{cc}
{\epsfig{file=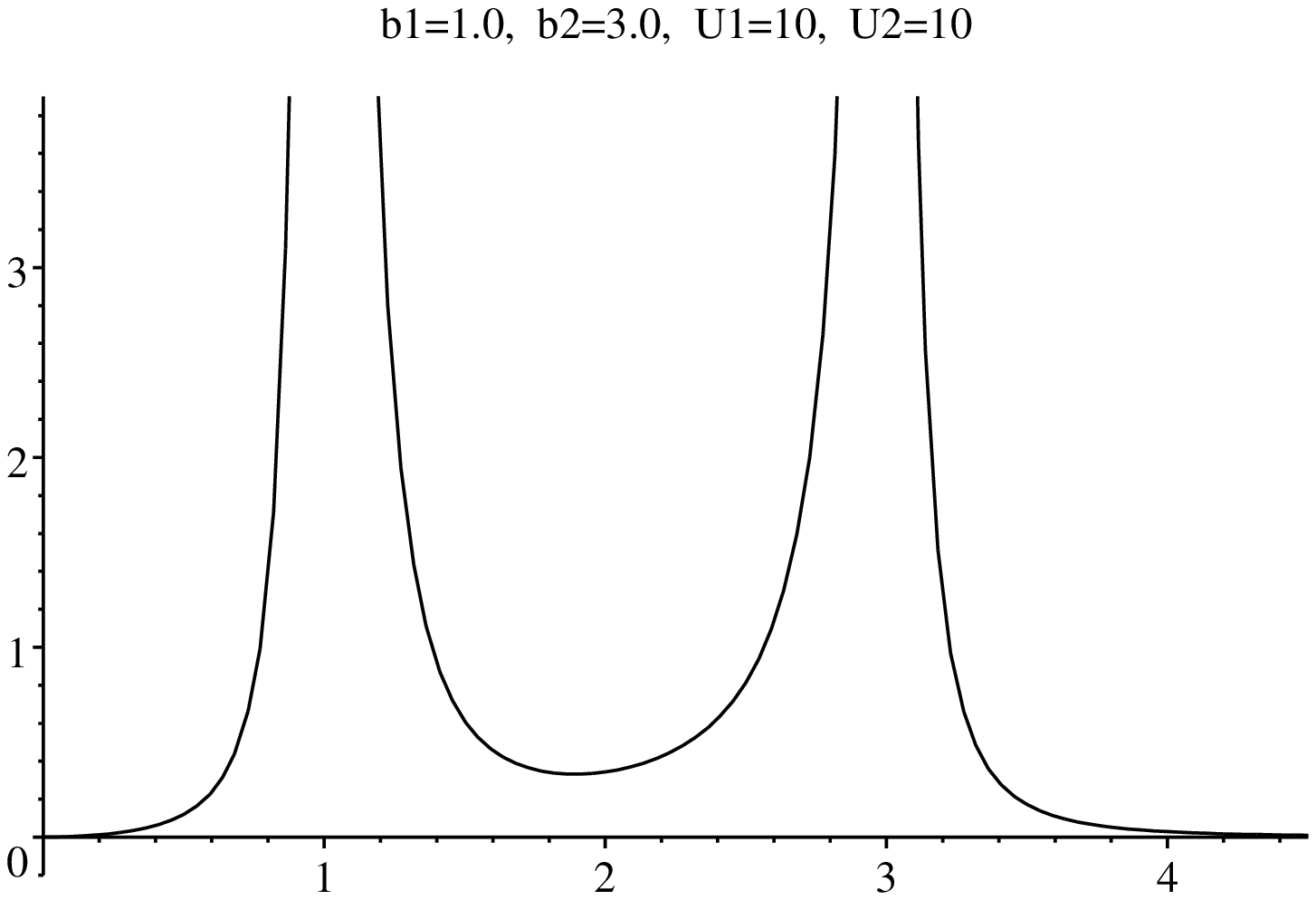, width=7cm}} &
{\epsfig{file=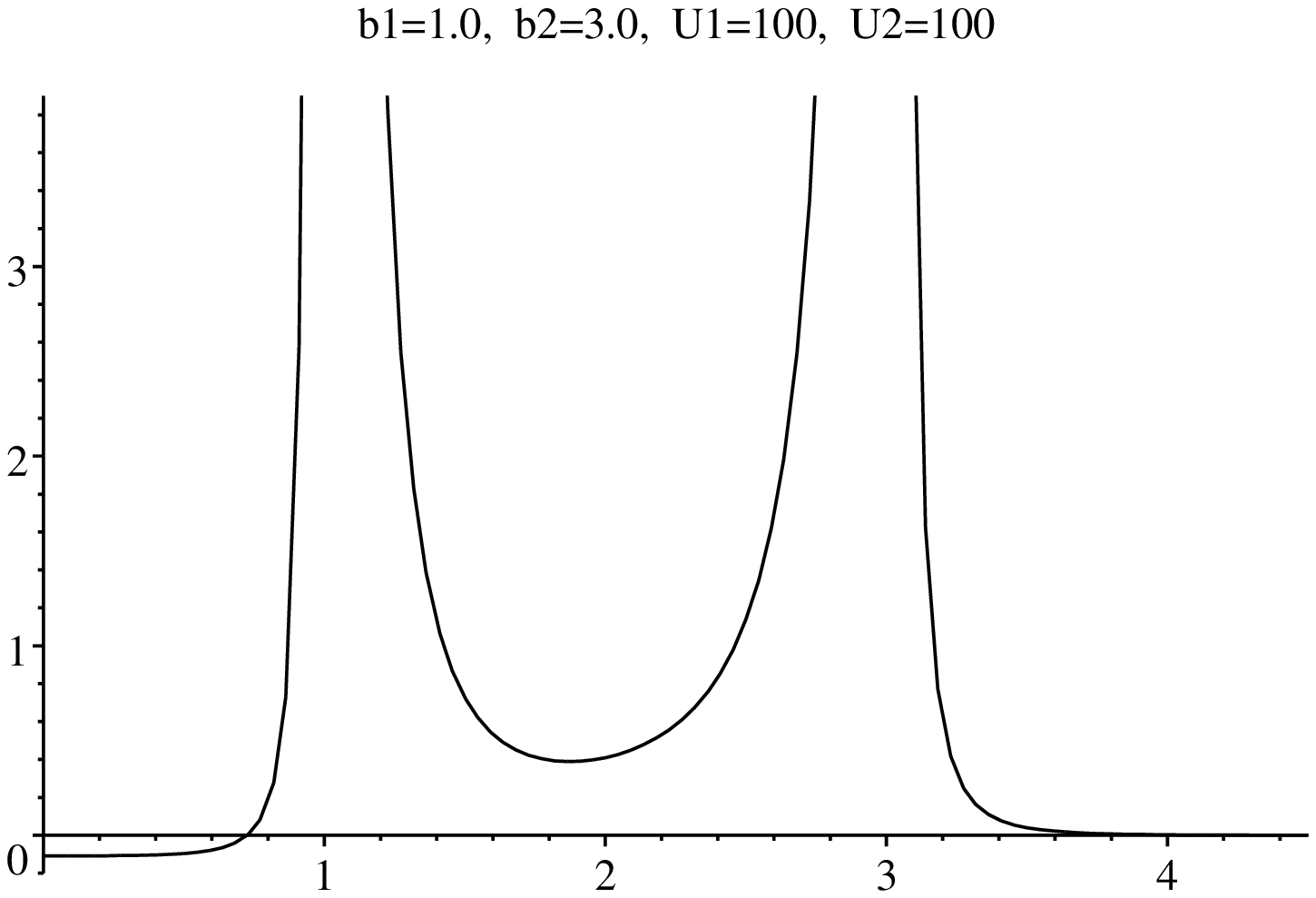, width=7cm}} \\
{\epsfig{file=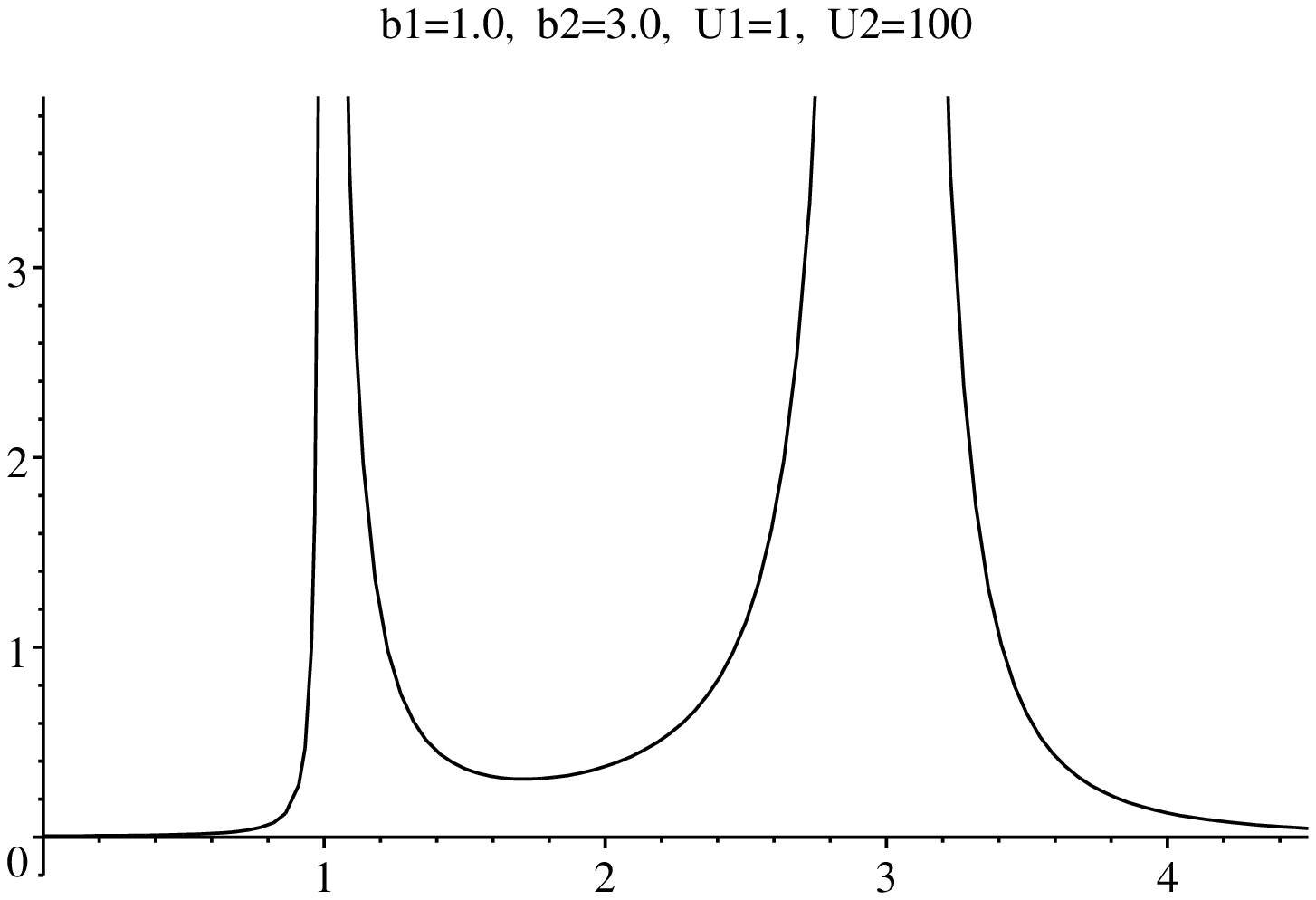, width=7cm}} &
{\epsfig{file=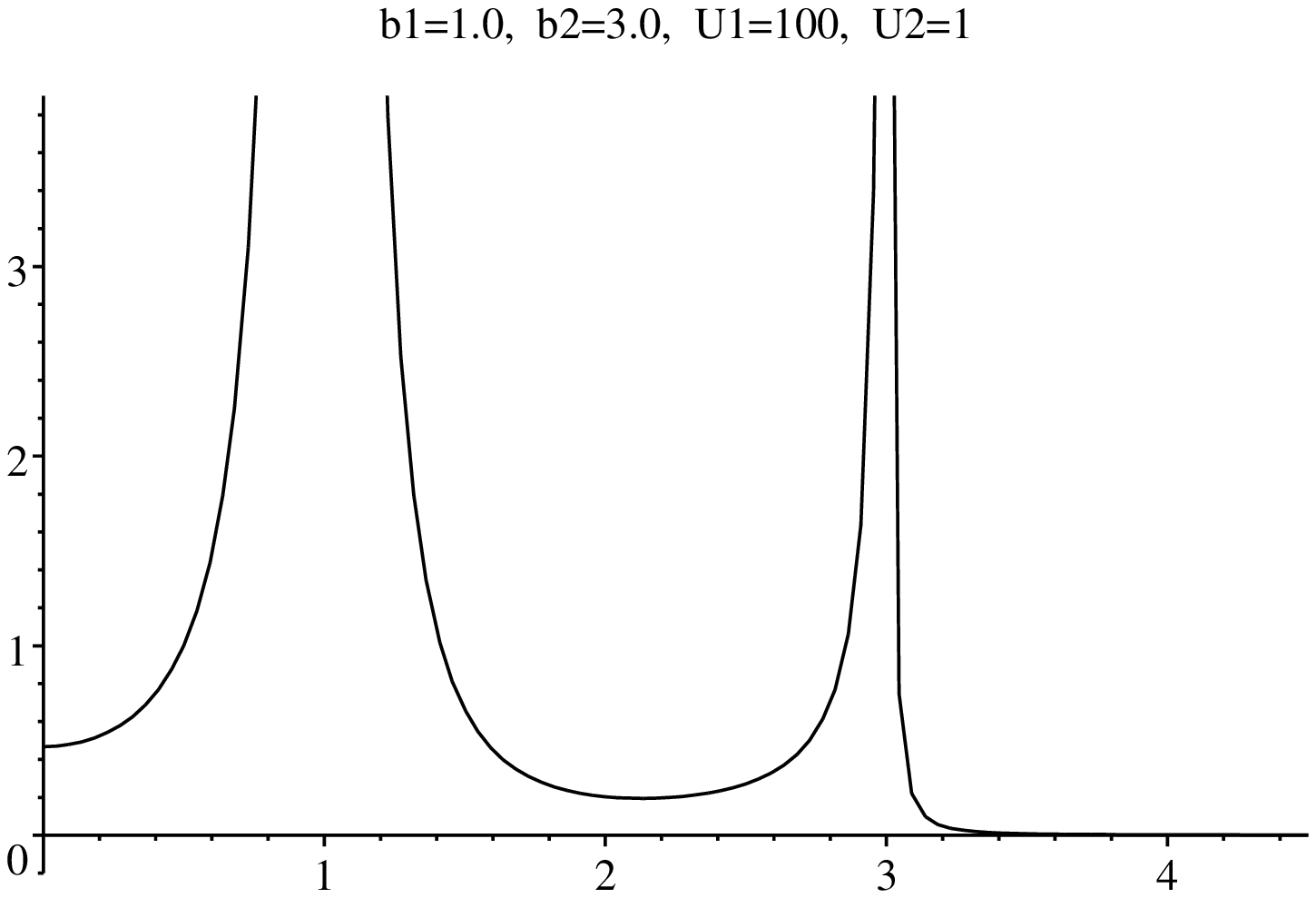, width=7cm}}
\end{tabular}
\caption[f6]
{$-4\pi^2 b^2 \langle \hat{\varphi}^2\rangle^{\ind{ren}}$ in
4-dimensional spacetime as the function of $\rho/b$ for different values
of the parameter $U_1$ and $U_2$.}
\label{f6}
\end{figure} 

To illustrate  a typical  behavior of  $\langle
\hat{\varphi}^2\rangle^{\ind{ren}}$ for the two-mirror problem, we plot
in Figure~6  the value of $-4\pi^2 b^2 \langle
\hat{\varphi}^2\rangle^{\ind{ren}}$ in 4-dimensional spacetime as a
function of $\rho$ for different values of the parameter $U_1$ and
$U_2$, and for $b_1=1$ and $b_2=3$.

\section{Stress-Energy Tensor}\label{s5}
\setcounter{equation}0

\subsection{General properties}

For both  problems $A$ and $B$, the boundary  conditions are invariant
under the group of rotations $O(d+1)$ which is the space symmetry. As a
result of  this symmetry, the renormalized stress-energy tensor has the
structure
\be
\langle\hat{T}_{\mu}^{\nu}\rangle ^{\ind{ren}} = \mbox{diag}\,(\epsilon ,p,
\ldots\,p)\,, \n{5.1}
\ee
where $\epsilon$ and $p$ are functions of $\rho$. The conservation law
\be
\langle\hat{T}_{\mu}^{\nu}\rangle ^{\ind{ren}}_{\ \ ;\nu} = 0 \n{5.2}
\ee
implies
\be
\frac{d\epsilon}{d\rho}
= -\,\frac{d}{\rho}\,(\epsilon - p)\,. \n{5.4}
\ee

For the conformal invariant theory (\ref{2.3})--- (\ref{2.4}) we have
$\langle\hat{T}_{\mu}^{\mu}\rangle ^{\ind{ren}}=\epsilon + d\cdot p=0$,
while the conservation law (\ref{5.4}) gives
\be
\epsilon  = \frac{\epsilon_0}{\rho^{d+1}}\,, \quad p = -\, \frac
{\epsilon_0}{d\,\rho^{d+1}}\,. \n{5.6}
\ee
In other words, the renormalized stress-energy tensor for the
conformal  invariant theory is uniquely determined by the symmetry and 
the  conservation law up to one constant. In a more general case, it is
sufficient to  determine only one function of one variable $\rho$, say
$\epsilon - p$.

Using equations (\ref{2.6}) and (\ref{2.7}), we get
\be
\epsilon  - p = \lim_{x'\rightarrow x}\left[ {\cal R} +{1\over
\rho^2} {\cal N} 
\right] G^{\ind{ren}}(x,x')\,. \n{5.7}
\ee
Here
\be \n{5.8}
{\cal R}= (1-2\xi )\partial_{\rho}\,
\partial_{\rho'} - \xi\left(\partial_{\rho}^2 + \partial_{\rho'}^2\right)
+{\xi\over \rho}(\partial_{\rho} +
\partial_{\rho'})\, ,
\ee
\be \n{5.9}
{\cal N}= -\,(1-2\xi )\,\partial_{\eta}\,\partial_{\eta'} +\xi
\left[\partial_{\eta}^2 + \partial_{\eta'}^2 \right]\, .
\ee
(The last term in the right-hand-side of (\ref{5.8}) arises because in
the spherical coordinates $\Gamma_{\eta\eta}^{\rho}= -\rho \ne 0$.)
We use now this relation to obtain the renormalized stress-energy tensors 
for our problems.

\subsection{Problem $A$}

Using expressions (\ref{3.40}), (\ref{3.41}), and (\ref{5.7}), we can write
\be
\epsilon - p = -\,\frac{U_0\,\eta_d}{
b^{d-1}}\left[\left({\cal R} + 
\frac{1}{\rho^2}\,{\cal N}\right)Q_{\pm}(\rho ,\rho',
\gamma )\right]_{ \begin{array}{l}
\rho' =\rho\,  \\  \gamma =0\,   
\end{array} }\,, \n{5.10}
\ee
where
\be
Q_{\pm}(\rho ,\rho',\gamma ) = \sum_{L=0}^{\infty}\frac{1}{(L+\beta_d+U_0)}\,
\left(\frac{\rho\rho'}{b^2}\right)^{L_{\pm}}\,C_L^{(d-1)/2}(\cos \gamma )\,.
 \n{5.11}
\ee
It is easy to verify that
\be
\left[{\cal R}\left(\rho\rho'\right)^{L_{\pm}}\right]_{\rho =\rho' } 
= \left[(1-4\xi )\,L_{\pm}^2 + 4\xi L_{\pm}\right]
\rho^{2(L_{\pm}-1)} \,, \n{5.12}
\ee
\be
\left[{\cal N}\,C_L^{(d-1)/2}(\cos \gamma )\right]_{\gamma =0 } 
=  \frac{L_+ \,L_-}{d}\,C_L^{(d-1)/2}(1) \,.\hspace{1.9cm}  \n{5.13}
\ee
In order to obtain (\ref{5.13}), we used (\ref{3.29}) and the following property 
of hypergeometric functions
\be
\left.\frac{d\ }{dz}\,F(a,b;c;z)\right|_{z =0 } =  \frac{ab}{c} \,.\n{5.14}
\ee

Using relations (\ref{5.12}), (\ref{5.13}), and (\ref{4.2}), we get
\be
\epsilon - p = -\,
\frac{U_0\,\eta_d}{b^{d-1}\,\rho^2 }\sum_{L=0}^{\infty}
\frac{{\cal A}_{\pm}^{\xi}}{(L+\beta_d+U_0)}
\,\frac{(L+d-2)!}{L!\,(d-2)!}
\left(\frac{\rho^2}{b^2}\right)^{L_{\pm}}\,, \n{5.15}
\ee
where
\be
{\cal A}^{\pm}_{L,\xi} = (1-4\xi )\,L_{\pm}^2 + 4\xi L_{\pm} +
\frac{L_+\,L_-}{d}
=L_{\pm}\, (L_{\pm}-1)\, ({d-1\over d}-4\xi)
\,. \n{5.16}
\ee
The signs $+$ and $-$ correspond to the internal $(\rho < b)$ and external 
$(\rho > b)$ problems, respectively.
Using relation (\ref{5.4}), we obtain the following expression for
$\epsilon$
\be
\epsilon = 
\frac{U_0\,\eta_d \, d}{2\, b^{d-1}\,\rho^2 }\sum_{L=0}^{\infty}
\frac{{\cal A}_{\pm}^{\xi}}{(L_{\pm}-1)\, (L+\beta_d+U_0)}
\,\frac{(L+d-2)!}{L!\,(d-2)!}
\left(\frac{\rho^2}{b^2}\right)^{L_{\pm}}\,, \n{5.15a}
\ee
It is easy to see that first two terms (with $L=0$ and $L=1$) (\ref{5.15})   
for the inner problem vanish. Hence, in the
general case, $\epsilon -p \sim \rho^2$ for small $\rho$, while
$\epsilon$ remains finite at $\rho=0$.

Equation (\ref{5.16}) shows that, for a special case of conformal
theory, when $\xi = \xi_d = \frac{d-1}{4d}$, the coefficients ${\cal
A}^{\pm}_{L,\xi_d}$ vanish. For this reason, $\epsilon = p$, and hence
$\epsilon = p = 0.$ This result  is similar to what happens in the 
case of an ideally reflecting mirror considered in \cite{FrSe:79}.

When $\xi = 0$, we have
\be
{\cal A}^{\pm}_{L,\xi=0} = \frac{d-1}{d}\,L_{\pm}( L_{\pm} -1)\,. \n{5.20}
\ee
Let us consider the internal $(\rho <b)$ problem first. In this case,
\be
{\cal A}^+_{L,\xi=0} = \frac{(d-1)L(L-1)}{d}\,. \n{5.21}
\ee
Substituting ${\cal A}^+_{L,\xi=0}$ into (\ref{5.15}), we find for $\rho
<b$
\be
\epsilon - p = - {\cal B}\, U_0\,\frac{\rho^2}{b^{d+3}}\,
F^{(d+2)}\left((\rho/b)^2,\,U_0+(d+3)/2\right), 
\n{5.22}
\ee
where $F^{(d)}$ is the function determined by relation (\ref{4.6}), and
\be\n{5.23}
{\cal B}\, = \eta_d\, (d-1)^2 \, .
\ee

Similarly, for the external $(\rho >b)$ problem,
\be
{\cal A}^-_{L,\xi=0} = \frac{(d-1)(L+d)(L+d-1)}{d}\,. \n{5.24}
\ee
Substituting this expression into (\ref{5.15}), we get
\be
\epsilon - p = -\,{\cal B}\, U_0\,\frac{b^{d-1}}{\rho^{2d}}\,
F^{d+2}\left((b/\rho )^2,\,U_0+(d-1)/2\right). 
\n{5.25}
\ee

Equation (\ref{5.4}) allows us  to obtain $\epsilon$ (and hence  $p$). 
Using  relations (\ref{A.8}) and (\ref{A.9}) we get
\be\n{5.28}
\epsilon = {{\cal B}\, U_0\over 2 b^{d+1}}\,
\left\{
\begin{array}{ll}
\displaystyle{
F^{(d+1)}(\left((\rho/b)^2,\,U_0+(d+1)/2\right))
}\, , 
& \rho < b\,,
 \\ 
\  & \\
\displaystyle{
-\left({b\over
\rho}\right)^{2d}
F^{(d+1)}(\left((b/\rho)^2,\,U_0+(d-1)/2\right))
}\, , 
& \rho > b\,.
\end{array}
\right.
\ee

For an ideal reflecting mirror, $U_0=\infty$, the expressions for
$\epsilon-p$ and $\epsilon$ are greatly simplified.  Using  relation
(\ref{A.7}) we get
\be\n{5.30}
\epsilon = \pm {{\cal B}\, \over 2}\,
{b^{d-1} \over \left| b^2 -\rho^2\right|^d}\, ,
\ee
\be\n{5.31}
\epsilon-p = -{\cal B}\,b^{d-1}\, {\rho^2 \over \left|
b^2-\rho^2\right|^{d-1}}\, .
\ee
The signs $\pm$ in (\ref{5.30}) correspond to the inner and outer problem,
respectively. For $d=3$, expressions (\ref{5.30}) -- (\ref{5.31})
reproduce the results obtained in \cite{FrSe:79}.

\subsection{Problem B}

\subsubsection{General formulas}

Calculations of $\epsilon -p$ and $\epsilon$ for problem B
are similar to the calculations for the problem $A$, but are much more
involved. Besides relations (\ref{5.12}) and (\ref{5.13}), we use also
the relations
\be
\left[ {\cal R}\left( \rho^{L_{\pm}}\,
{\rho'}^{L_{\mp}}\right)\right]_{\rho'=\rho}= {1\over \rho^{d+1}}[L_+
L_- -\xi (d^2-1)]\, .
\ee
For this section, we only list the final results.

\medskip

\noindent
{\bf Inner region} $\rho < b_1$: 
\be\n{5.32}
\epsilon -p = -{\eta_d \over  \rho^2}\, 
\sum_{L=0}^{\infty} {(L+\beta_d)\, (L+d-2)!\over L!\, (d-2)!} \,
{\cal A}^{+}_{L,\xi}\,  {\cal R}_L^+ (\rho)\, ,
\ee
\be\n{5.33}
\epsilon = {\eta_d\, d \over 2 \rho^2}\, 
\sum_{L=0}^{\infty} {(L+\beta_d)\, (L+d-2)!\over L!\, (d-2)!\, (L-1)} \,
{\cal A}^{+}_{L,\xi}\,  {\cal R}_L^+ (\rho)\, ,
\ee
\be \n{5.34}
{\cal R}_L^+(\rho)= {1\over b_1^{d-1}}\, \left( {\rho \over b_1}
\right)^{2L}\, {U_1 +U_2 \gamma_L +{\displaystyle{U_1 U_2 \over
L+\beta_d}}\,
(1-\gamma_L)\over
(L + \beta_d + U_1)(L + \beta_d + U_2) - U_1 U_2
\gamma_L 
}
\, .
\ee

\medskip

\noindent
{\bf Outer region} $\rho > b_2$: 

\be\n{5.36}
\epsilon -p =-{\eta_d b_2^{d-1}\over  \rho^{2d}}\, 
\sum_{L=0}^{\infty} {(L+\beta_d)\, (L+d-2)!\over L!\, (d-2)!\, } \,
{\cal A}^{-}_{L,\xi}\,  {\cal R}_L^-(\rho)\, ,
\ee
\be\n{5.37}
\epsilon =-{\eta_d b_2^{d-1}\, d\over  2 \rho^{2d}}\, 
\sum_{L=0}^{\infty} {(L+\beta_d)\, (L+d-2)!\over L!\, (d-2)!\, (L+d)} \,
{\cal A}^{-}_{L,\xi}\,  {\cal R}_L^-(\rho)\, ,
\ee
\be\n{5.38}
{\cal R}_L^-(\rho)= \left({b_2 \over \rho}\right)^{2L}\, 
\, {U_1 \gamma_L +U_2 +{\displaystyle{U_1 U_2 \over L+\beta_d}}\,
(1-\gamma_L)\over
(L + \beta_d + U_1)(L + \beta_d + U_2) - U_1 U_2
\gamma_L
}
\, .
\ee

\medskip

\noindent
{\bf Intermediate region}  $ b_2 > \rho >   b_1$:

\be\n{5.39}
\epsilon -p =-{\eta_d \over  \rho^{d+1}}\, 
\sum_{L=0}^{\infty} {(L+d-2)!\over L!\, (d-2)!} \,
\left[
{\cal A}^{-}_{L,\xi}\,  \tilde{{\cal R}}_L^-(\rho)+
{\cal A}^{+}_{L,\xi}\,  \tilde{{\cal R}}_L^+(\rho)+
{\cal A}^{0}_{L,\xi}\,  \tilde{{\cal R}}_L^0
\right]
\, .
\ee
\be\n{5.39a}
\epsilon ={\eta_d \, d\over  2\rho^{d+1}}\, 
\sum_{L=0}^{\infty} {(L+d-2)!\over L!\, (d-2)!} \,
\left[
{{\cal A}^{-}_{L,\xi}\over (L_- -1)}\,  \tilde{{\cal R}}_L^-(\rho)+
{{\cal A}^{+}_{L,\xi}\over (L_+ -1)}\,  \tilde{{\cal R}}_L^+(\rho)
-{2\over d+1}\, {\cal A}^{0}_{L,\xi}\,  \tilde{{\cal R}}_L^0
\right]
\, .
\ee

\bigskip

The coefficients ${\cal A}^{\pm}_{L,\xi}$ in relations (\ref{5.32}),
(\ref{5.36}) and (\ref{5.39}) are given by equation (\ref{5.16}), and 
\be\n{5.40}
{\cal A}^{0}_{L,\xi}={d+1\over d}\, L_+\, L_- -\xi (d^2-1)=
-\left[{d+1\over d}\,L\, (L+d-1)+ \xi\, (d^2-1)\right]\, .
\ee
We also use the notations
\be\n{5.41}
\tilde{{\cal R}}_L^-(\rho)={U_1\, (L+\beta_d +U_2)\over 
(L + \beta_d + U_1)(L + \beta_d + U_2) - U_1 U_2\gamma_L }
\left( b_1\over \rho\right)^{2L+d-1}\, ,
\ee
\be\n{5.42}
\tilde{{\cal R}}_L^+(\rho)={U_2\, (L+\beta_d +U_1)\over 
(L + \beta_d + U_1)(L + \beta_d + U_2) - U_1 U_2\gamma_L }
\left( \rho\over b_2\right)^{2L+d-1}\, ,
\ee
\be\n{5.43}
\tilde{{\cal R}}_L^0 = {2\, U_1\, U_2\, \gamma_L\over 
(L + \beta_d + U_1)(L + \beta_d + U_2) - U_1 U_2\gamma_L } \, .
\ee
Expressions for $\epsilon$ in each of the three regions can be easily
obtained by integrating relation (\ref{5.4}).

\subsubsection{Conformally invariant theory}

Equation (\ref{5.16}) implies that ${\cal A}^{\pm}_{L,\xi}=0$ for the
conformal invariant theory, when $\xi=\xi_d=(d-1)/4d$. 
Equations (\ref{5.33}), (\ref{5.34}), (\ref{5.36}), and (\ref{5.37})
show that
$\epsilon=p=0 $ both in the inner and outer regions. In the
intermediate region, we have
\be\n{5.44}
\epsilon -p ={C\over \rho^{d+1}}\, ,
\ee
where
\be\n{5.45}
C= 2 U_1\, U_2\, \eta_d \, {d+1\over d} \, 
\sum_{L=0}^{\infty} {(L+d-2)!\over L!\, (d-2)!}\, 
{{\displaystyle\gamma_L\left[L(L+d-1)+  {(d-1)^2\over 4}\right]
\over 
(L + \beta_d + U_1)(L + \beta_d + U_2) - U_1 U_2\gamma_L }} \, .
\ee
Using relation $\epsilon=-d\, p$ we get
\be\n{5.46}
\epsilon = {d\over d+1}{C\over \rho^{d+1}}\, .
\ee
It is easy to verify that in the limit of ideally reflecting mirrors
($U_1, U_2 \rightarrow \infty$) in 4-dimensional space ($d=3$),
relations (\ref{5.44}) -- (\ref{5.46})  correctly reproduce the results of
\cite{FrSe:79}. 

\subsubsection{Two-dimensional case}

We now make several remarks concerning $\langle
\hat{T}^{\mu}_{\nu}\rangle^{\ind{ren}}$ calculations in a 2-dimensional
spacetime\footnote{We shall not consider $\langle
\hat{\varphi}^2\rangle^{\ind{ren}}$ because of the infrared problem it
 is not a well-behaved quantity. In particular, $\langle
\hat{\varphi}^2\rangle^{\ind{ren}}$ is logarithmically
divergent at infinity for any value of the mirror potential.}. 
Since $\xi_{d=1}=0$, conformal and canonical stress-energy tensors are
identical in $2$-dimensional spacetime. We restrict ourselves by
discussing only this special case and put $\xi=0$. In the
$2$-dimensional case, the spherical harmonics (which enter for example
into the expression (\ref{3.23}) for the Green function) are simply
functions $\exp(im\eta)/\sqrt{2\pi}$, and $(L,W)\equiv m= 0,\pm 1,\pm
2,\dots$ . For all modes with $m\ne 0,$ the general relation
(\ref{3.52}) can be easily verified for the $2$-dimensional case. (One
need only  to put $\beta_d=0$ there.)  The only problem is due to the
mode $m=0$. A general solution of the radial equation
(\ref{3.24}) for $m=0$ is logarithmically divergent either at $\rho=0$
or $\rho=\infty$. For this reason, in the general case there does not
exist a radial Green function ${\cal G}_{m=0}(\rho,\rho')$ which
remains finite at both boundaries. Fortunately, these zero modes
does not contribute to $\langle \hat{T}^{\mu}_{\nu}\rangle^{\ind{ren}}$
and the corresponding ambiguity is not important for our problem. The
calculations for the problem $B$  give  $\epsilon=p=0$ inside the inner
mirror and outside the outer mirror, while between the mirrors we have
\ba
\epsilon - p & = & - \frac{2}{\pi} \frac{U_1 \, U_2}{\rho^2}
 \sum_{m=1}^{\infty}
\frac{m \, \gamma_m}{\Omega_m}\, ,
\nonumber \\
\epsilon & = & - \frac{U_1 \, U_2}{\pi \rho^2} \sum_{m=1}^{\infty}
\frac{m \, \gamma_m}{\Omega_m}\, .
\ea
Here, $\gamma_m  = \left(b_1/b_2\right)^{2m}$ and
$\Omega_m  =  (m + U_1)(m + U_2) - U_1 \, U_2 \, \gamma_m$ .
Both quantities $\epsilon$ and $p$ evidently vanish if one of the
potentials $U_i$ vanishes. This is exactly the result which must be
expected for Problem $A$.

\section{Radiation From Mirrors}\label{s6}
\setcounter{equation}0

Let us now discuss the  properties of quantum radiation created by
spherical mirrors which expand with a constant acceleration (see
footnote 1). Our primary aim is to derive an expression for radiation 
at infinity produced by such a mirror. This problem can be solved by
making an analytical continuation of the results of calculations for
$\langle \hat{\varphi}^2\rangle^{\ind{ren}}$ and  $\langle
\hat{T}^{\mu}_{\nu}\rangle^{\ind{ren}}$ in the Euclidean space.

\subsection{$\langle \hat{\varphi}^2\rangle^{\ind{ren}}$ at ${\cal J}^+$}

First, let us study  $\langle \hat{\varphi}^2\rangle^{\ind{ren}}$
at  ${\cal J}^+$. We consider only Problem {\em B}, since the
result for  problem $A$ can be easily obtained by taking the limit
$U_1\rightarrow 0$. For Problem $B,$ in the exterior region we
have (cf. (\ref{4.22}) and (\ref{5.38}))
\be\n{6.1} 
\langle \hat{\varphi}^2\rangle^{\ind{ren}} (\rho)=-{\eta_d
b_2^{d-1}\over \rho^{2(d-1)}}
\sum_{L=0}^{\infty} {(L+\beta_d)\, (L+d-2)!\over L!\, (d-2)!\, } \,
{\cal R}_L^-(\rho)\, .
\ee
Using (\ref{3.7}), we can write
\be\n{6.2}
\rho^2 =R^2+X_0^2\, .
\ee
After performing a Wick's rotation $X_0 =iT,$ this quantity becomes
\be\n{6.3}
\rho^2 =R^2 -T^2\, .
\ee
By letting $u=T-R$ be a retarded time, we have at large $R$
\be\n{6.4}
\rho^2 \approx -2uR\, .
\ee
For this definition of $u$, the value $u=0$ corresponds to the moment of the
retarded time when both expanding mirrors reach ${\cal J}^+$. For a part
of ${\cal J}^+$ lying to the past of this moment 
(that is outside the mirror) we have
$u<0$.
The leading contribution at ${\cal J}^+$ is given by the term $L=0$ in
(\ref{6.1}). Hence we have 
\be
\n{6.5} 
\langle \hat{\varphi}^2\rangle^{\ind{ren}} (u)\approx -{\eta_d \over
R^{d-1}}
{b_2^{d-1}\over (-2u)^{d-1}}
{d-1\over 2} {\cal R}^-_0 \, ,
\ee
\be \n{6.6} 
{\cal R}^-_0 = {U_1 \gamma_0 +U_2 +{\displaystyle{U_1 U_2 \over \beta_d}}\,
(1-\gamma_0)\over
(\beta_d + U_1)(\beta_d + U_2) - U_1 U_2 \gamma_0
}
\, ,
\ee
where $\gamma_0 =(b_1/b_2)^{d-1}$.

\subsection{$\langle \hat{T}^{\mu}_{\nu}\rangle^{\ind{ren}}$ at ${\cal J}^+$}

Our starting point is the general expression for 
$\langle \hat{T}^{\mu}_{\nu}\rangle^{\ind{ren}}$, which in the Euclidean
space under consideration takes the form
\be\n{6.7}
T^{\mu}_{\nu}\equiv \langle \hat{T}^{\mu}_{\nu}\rangle^{\ind{ren}}=
(\epsilon -p)\, \delta^{(\rho)}_{\nu}\,
\delta_{(\rho)}^{\mu}+p\delta^{\mu}_{\nu}\, .
\ee
Here $\delta^{(\rho)}_{\nu}$ is the Kronecker symbol which is
non-vanishing
only when $\nu$ is the index corresponding to the co-ordinate $\rho$.
(No summation over $\rho$!)
Using  expression (\ref{6.7}), we get in ($X_0, R$) coordinates
\begin{eqnarray} \n{6.8}
T_{X_0 X_0} & = & {1\over \rho^2}\left( X_0^2\epsilon +R^2 p \right)\, ,
\nonumber \\
T_{X_0 R} & = & {1\over \rho^2}\, X_0\, R\, (\epsilon- p)\, ,\\
T_{RR} & = & {1\over \rho^2}\left( R^2\epsilon +X_0^2 p \right)\, . \nonumber
\end{eqnarray}
By making a Wick's rotation $X_0=iT,$ we get
\begin{eqnarray} \n{6.9}
T_{TT} & = & {1\over \rho^2}\left( T^2\epsilon -R^2 p \right)\, ,
\nonumber \\
T_{T R} & = & -{1\over \rho^2}\, T\, R\, (\epsilon- p)\, ,\\
T_{RR} & = & {1\over \rho^2}\left( R^2\epsilon -T^2 p \right)\, . \nonumber
\end{eqnarray}
For fixed $u=T-R$ and $R\rightarrow \infty$ we find
\be\n{6.10}
T_{TT}\approx -T_{T R} \approx T_{RR}  \approx {R^2\over \rho^2}
(\epsilon -p)\, .
\ee
For large $R$, the leading contribution to $\epsilon -p$ is given by the
$L=0$ term in (\ref{5.36}). Thus, we have
\be\n{6.11}
\epsilon -p\approx -{\eta_d b_2^{d-1}\over \rho^{2d}}{d-1\over 2} {\cal
A}^-_{0,\xi} {\cal R}^-_0\, ,
\ee
where ${\cal R}^-_0$ is given by (\ref{6.6}) and
\be\n{6.12}
{\cal A}^-_{0,\xi} =(d-1)\, \left(d-1-4\xi\, d\right)\, .
\ee

Combining these results, we get for $T_{\mu\nu}$ in the asymptotic region
the following expression
\be\n{6.13}
T_{\mu\nu}={\dot{E}\over {\cal S}_{d-1}}\, l_{\mu}\, l_{\nu}\, ,
\ee
where $l_{\mu}=u_{,\mu}$, $u=T-R$ is retarded time, 
\be\n{6.14}
\dot{E}\equiv {dE\over du}=-{1\over 4\pi}\, B\left({d-1\over 2}, {1\over
2}\right)\,
{b_2^{d-1}\over (-2u)^{d+1}}\, (d-1)^2\, (d-1-4\xi\, d)\, {\cal R}^-_0\,
,
\ee
and
\be\n{6.15}
{\cal S}_{d-1}=R^{d-1}\, {\cal V}_{d-1}\,
\ee
is the surface area of a $(d-1)$-dimensional sphere of radius $R$. The 
function
$B(z,w)$ in relation (\ref{6.14}) is the Beta function
$B(z,w)=\Gamma(z)\Gamma(w)/\Gamma(z+w)$.

For an ideally reflecting external mirror, $U_2 =\infty$, ${\cal
R}^-_0=2/(d-1)$, so that for the radiation of such a mirror in
4-dimensional spacetime ($d=3$) we have
\be\n{6.16}
{dE\over du}=-{b_2^2\over 4\pi\, u^4}(1-6\xi)\, .
\ee
For $\xi=0$, this result reproduces the result obtained in
\cite{FrSe:79}.

By making the substitution $u\rightarrow v=T+R$ in relations  (\ref{6.5})
and (\ref{6.13})-- (\ref{6.14}) we easily obtain expressions  for
$\langle \hat{\varphi}^2\rangle^{\ind{ren}}$  and $T_{\mu\nu}$ at ${\cal
J}^-$. 

In a similar way it is easy to obtain expressions for  $\langle
\hat{\varphi}^2\rangle^{\ind{ren}}$  and $T_{\mu\nu}$ in the region
located inside the inner mirror or in the region between the mirrors (for
problem B). We do not reproduce here the corresponding results. We
only mention that both $\langle \hat{\varphi}^2\rangle^{\ind{ren}}$
and  $T_{\mu\nu}$ remain finite on the null cone $N: \, \, R^2=T^2$ for
any finite value of time $T$. This property directly follows from the
finiteness of  $\langle \hat{\varphi}^2\rangle^{\ind{ren}}$ and 
$T_{\mu\nu}$ at the origin $\rho =0$ in the Euclidean formulation. The
null cone $N$ plays the role of the event horizon for an observer
moving on the mirror surface. Regularity of physical observables on the
horizon differs the case of moving mirrors from the case of the black
hole (see e.g. discussion in \cite{Pare:93,MaPa:97}).

\section{Conclusion}

In this paper, we studied the quantum effects generated by spherical
partially transparent mirrors expanding with a constant acceleration
in a $D$-dimensional flat spacetime. We consider a scalar massless
field with an arbitrary parameter of non-minimal coupling $\xi$. The
choice of this parameter does not affect the field equation and the
expectation value of $\langle \hat{\varphi}^2 \rangle^{\ind{ren}}$  but
results in different expressions for the stress-energy tensor. A
partially transparent mirror is modeled by a $\delta$-like potential
in the field equation. Two special models were studied: Model $A$ with
a single mirror which expands with a constant acceleration so that its
worldsheet is described by the equation $R^2-T^2=b^2$, where $b^{-1}$ is
the value of the acceleration, and Model $B$ with a couple of
concentric expanding mirrors.  

We demonstrated that the leading terms of $\langle \hat{\varphi}^2
\rangle^{\ind{ren}}$ and $\langle \hat{T}^{\mu}_{\nu}
\rangle^{\ind{ren}}$ at ${\cal J}^+$  has quite a simple form, 
(\ref{6.5}) and (\ref{6.13})--(\ref{6.14}), respectively. Both
expressions infinitely grow at
$u=0$, the moment of the retarded time when the mirrors reach ${\cal
J}^+$. The same divergence takes place at the moment $v=0$ of the
advanced time when the mirrors start their motion from  ${\cal J}^-$. Both
of these divergences are evidently connected with the adopted
idealization of the problem: an infinite time of the accelerated motion
and infinite sharpness of the mirrors.
It should be also emphasized that since we used the Euclidean approach
and obtained $\langle \hat{\varphi}^2
\rangle^{\ind{ren}}$ and $\langle \hat{T}^{\mu}_{\nu}
\rangle^{\ind{ren}}$ by using the Wick's rotation from their Euclidean
values, the corresponding quantities in the physical spacetime are given
for a special choice of state of the quantum field, namely the state
which is invariant under time reflection $T\rightarrow -T$ 
(for more details of the
properties of such states, see \cite{FrSe:79} and \cite{FrSe:80}). 

It is interesting to note that in the model $B$ for the conformally
invariant case (that is, for a special choice of the coupling
$\xi=(d-1)/(4d)$), the stress-energy tensor identically vanishes both
inside the inner mirror and outside the external mirror. For the
conformal field in Model
$A$ this occurs everywhere. As indicated earlier (see
\cite{FrSe:79} and \cite{FrSe:80}), this is a direct consequence of the
conformal invariance of the models.  A more general discussion of the
properties of the vacuum in Minkowski spacetime under conformal
transformations can be found in \cite{JaRe:95} and \cite{JaRe:97}.

\bigskip

\vspace{12pt}
{\bf Acknowledgments}:\ \  This work was  partly supported  by  the
Natural Sciences and Engineering Research Council of Canada. One of the
authors (V.F.) is grateful to the Killam Trust for its financial
support.

\bigskip
\newpage

\appendix
\section{Function $F^{(d)}(z,\beta)$ and its properties}
\label{1+3_expansions}
\setcounter{equation}0

In this Appendix we collect some useful formulae connected with 
properties of function $F^{(d)}(z,\beta)$. This function is defined by
the series
\be\n{A.1}
F^{(d)}(z,\beta )  = \sum_{L=0}^{\infty} \frac{(L+d-2)!}{L!\,(d-2)!\,(L+\beta )}
\,z^{L} \,.
\ee
It also  allows the following representation (\cite 
{PrBrMa:86}, v.1, equation 5.2.11.15):
\be
F^{(d)}(z,\beta )  = \int_0^{\infty}\frac{e^{-\beta t}\,dt}{(1-ze^{-t})^{d-1}}\,. 
\n{A.2}
\ee   
Using this representation it is easy to see that
\be
F^{(d)}(0,\beta )  = \frac{1}{\beta }\,. \n{A.3}
\ee   

The function $F^{(d)}(z,\beta)$ is defined for $z<1,$ where the series
(\ref{A.1}) is convergent. Near $z=1,$ it has a divergence. Since this
divergence is connected with the behavior of the series at large $L$,
one can estimate the leading divergence by putting $\beta=1$ in
(\ref{A.1}). Using the relation (\cite{PrBrMa:86}, v.1 relation
5.2.11.1)
\[
\sum_{L=0}^{\infty} {(L+d-2)!\over (L+1)! (d-2)!} z^L=
{1\over (1-z)^{d-2}} \sum_{k=0}^{d-3}{(d-3)!\over (k+1)! (d-3-k)!}(-z)^{k}
\]
\be\n{A.4}
\qquad = {1-(1-z)^{d-2} \over (d-2) z (1-z)^{d-2}}\, 
\ee
we can conclude that the leading divergence of $F^{(d)}$ near $z=1$ is of
the form
\be\n{A.5}
F^{(d)}(z,\beta)\sim {1\over (d-2)(1-z)^{d-2}}\, .
\ee
In the main text, we also use  the relation
\be\n{A.6}
\frac{1}{(1-x)^{d-1}}  = \sum_{L=0}^{\infty} \frac{(L+d-2)!}{L!\,(d-2)!}\,x^L 
\, .
\ee
This relation  allows one to show that
\be\n{A.7}
\lim_{\beta\rightarrow \infty}\left[\beta F^{(d)}(z,\beta)\right]={1\over
(1-z)^{d-1}}\, .
\ee
Using integral representation (\ref{A.2}) one can verify that 
the following two relations are valid for functions $F^{(d)}(z,\beta)$ 
\be\n{A.8}
\int_0^{z} dz\, F^{(d)}(z,\beta)\, =
{1\over d-2}\, F^{(d-1)}(z,\beta-1)\, ,
\ee
\be\n{A.9}
\int_0^{z} dz\, z^{d-2}\, F^{(d+1)}(z,\beta)\, =
{1\over d-1}\, z^{d-1}\, F^{(d)}(z,\beta)\, .
\ee


\newpage






\bigskip

\begin{thebibliography}{9}
\bibitem{Casi:48} H. B. G. Casimir,  Proc. K. Ned. akad. Wet. 
{\bf 51}, 793 (1948).
\bibitem{Mill:94} P. W. Milloni, {\em The Quantum Vacuum}, Academic
Press, New York, 1994.
\bibitem{Most:97} V. M. Mostepanenko,  {\em The Casimir Effect},
Cambridge Univ. Press, 1997.
\bibitem{Moor:70} G. T. Moore,  J. Math. Phys. {\bf 11}, 2679 (1970).
\bibitem{FuDa:76} S. A. Fulling, and P. C. W. Davies, Proc. Roy. Soc.
{\bf A 348}, 393 (1976).
\bibitem{DeWi:75} B. S. DeWitt, Phys. Rep. {\bf C 19}, 297 (1975).
\bibitem{BiDa:82} B. D. Birrel and P. C. W. Davies, {\em Quantum Fields
in Curved Spacetime}, Cambridge Univ. Press, 1982.
\bibitem{BoPeRo:84} M. Bordag, G. Petrov, and D. Robaschik.
 Sov. J. Nucl. Phys. {\bf 39}, 828 (1984).
\bibitem{BoPeRo:86} M. Bordag, G. Petrov, and D. Robaschik.
Sov. J. Nucl. Phys. {\bf 43}, 1034 (1986).
\bibitem{Calu:92} G. Calucci, J. Phys. A: Math. Gen. {\bf 25}, 3873
(1992).
\bibitem{Law:94} C. K. Law, Phys. Rev. {\bf A 49}, 433 (1994).
\bibitem{Dodo:95} V. V. Dodonov, Phys. Lett. {\bf A 207}, 126 (1995).
\bibitem{MeGi:96} O. Meplan and C. Gignoux, Phys. Rev. Lett. {\bf 76},
408 (1996).
\bibitem{LaJaRe:96} A. Lambrecht, M.-T. Jaekel, and S. Reynaud, Phys.
Rev. Lett. {\bf 77}, 615 (1996).
\bibitem{LaJaRe:98} A. Lambrecht, M.-T. Jaekel, and S. Reynaud,
Europhys. Lett. {\bf 43}, 147 (1998).
\bibitem{LaJaRe:98a} A. Lambrecht, M.-T. Jaekel, and S. Reynaud, {\em
Frequency Up-Converted Radiation from a Cavity Moving in Vacuum},
E-preprint quant-ph/9805044.
\bibitem{CoKa:98} R. Colestanian and M. Kardar, {\em Path Integral
Approach to the Dynamical Casimir Effect with Fluctuating Boundaries},
E-preprint quant-ph/9802017.
\bibitem{BaEb:93} G. Barton and C. Eberlein, Ann. Phys. (N.Y.) {\bf
227}, 222 (1993).
\bibitem{Bart:96} G. Barton, Ann. Phys. (N.Y.) {\bf
245}, 361 (1996).
\bibitem{BaNo:96} G. Barton and C. A. North, Ann. Phys. (N.Y.) {\bf
252}, 222 (1996).
\bibitem{CaDe:77} P. Candelas and S. Deutsch, Proc. R. Soc. {\bf A 354},
79 (1977).
\bibitem{FrSe:79}   V. P. Frolov and E. M. Serebriany,
Journ. Phys {\bf A 12}, 2415 (1979).
\bibitem{FrSe:80}   V. P. Frolov and E. M. Serebriany, 
Journ. Phys {\bf A 13}, 3205 (1980).
\bibitem{Cole:80} S. Coleman and F. De Luccia,  Phys.Rev. 
{\bf D 21}, 3305  (1980).
\bibitem{BeKuTk:87} V. A. Berezin, V. A. Kuzmin, and I. I. Tkachev, Phys. Rev.
{\bf D 36}, 2919 (1987).
\bibitem{BoKiVa:99}  M. Bordag, K. Kirsten, and D. Vassilevich,
Phys. Rev. {\bf D59}, 085011 (1999).
\bibitem{BoVa:99}  M. Bordag and D. Vassilevich, {\em Heat kernel expansion for 
semitransparent boundaries.} E-print hep-th/9907076.
\bibitem{ChMy:84}  A. Chodos, and  E. Myers, Ann. Phys. (N.Y.) {\bf 156},
412 (1984).
\bibitem{Ratr:85} B. Ratra,  Phys. Rev. {\bf D31} 31 (1985).
\bibitem{Higu:87} A. J. Higuchi,  Math. Phys. {\bf 28} 1553 (1987).
\bibitem{MTW} C. W. Misner, K. S. Thorne,  and J. A. Wheeler,  {\em
Gravitation} (W. H.~Freeman, San Francisco, 1973). 
\bibitem{PrBrMa:86} A. P. Prudnikov, Yu. A. Brychkov, and O. I.
Marichev, {\em Integrals and Series}, v.1. {\em Elementary Functions},
Gordon and Breach Publ., 1986.
\bibitem{Pare:93} R. Parentani, Class. Quant. Grav. {\bf 10}, 1409 (1993).
\bibitem{MaPa:97} S. Massar, R. Parentani, Phys.Rev.Lett. {\bf 78}, 4526
(1997).
\bibitem{JaRe:95}  M.-T. Jaekel and S. Reynaud, Quant. Semiclass. Opt.
{\bf 7}, 499 (1995).
\bibitem{JaRe:97}  M.-T. Jaekel and S. Reynaud, Rept. Prog. Phys. {\bf
60}, 863 (1997). 




\end{thebibliography}
\end{document}